\newcommand*{\ATLASLATEXPATH}{}
\documentclass[PAPER, atlasdraft=false, cernpreprint, texlive=2016, USenglish, english]{\ATLASLATEXPATH atlasdoc}
 
\usepackage[backend=biber]{\ATLASLATEXPATH atlaspackage}
\usepackage[backref=false]{\ATLASLATEXPATH atlasbiblatex}
 
\usepackage{\ATLASLATEXPATH atlasphysics}
 
\graphicspath{{logos/}{figures/}}
 
\usepackage{ANA-SUSY-2018-33-PAPER-defs}
 
\usepackage{adjustbox}
\usepackage{multirow,bigdelim}

\AtlasTitle{Search for long-lived, massive particles in events with a displaced vertex and a muon with large impact parameter in $pp$ collisions at $\sqrt{s} = 13$~TeV with the ATLAS detector}
 
\AtlasAbstract{
A search for long-lived particles decaying into hadrons and at least one muon is presented. The analysis selects events that pass a muon or missing-transverse-momentum trigger and contain a displaced muon track and a displaced vertex. The analyzed dataset of proton--proton collisions at $\sqrt{s} = 13$~TeV was collected with the ATLAS detector and corresponds to 136~fb$^{-1}$. The search employs dedicated reconstruction techniques that significantly increase the sensitivity to long-lived particle decays that occur in the ATLAS inner detector. Background estimates for Standard Model processes and instrumental effects are extracted from data. The observed event yields are compatible with those expected from background processes. The results are presented as limits at 95\% confidence level on model-independent cross sections for processes beyond the Standard Model, and interpreted as exclusion limits in scenarios with pair-production of long-lived top squarks that decay via a small $R$-parity-violating coupling into a quark and a muon. Top squarks with masses up to 1.7~TeV are excluded for a lifetime of 0.1~ns, and masses below 1.3~TeV are excluded for lifetimes between 0.01~ns and 30~ns.
}
 
\author{The ATLAS Collaboration}
 
\AtlasRefCode{SUSY-2018-33}
 
\PreprintIdNumber{CERN-EP-2019-219}

\AtlasCoverSupportingNote{Supporting Note}{https://cds.cern.ch/record/2633973/}
 
\AtlasCoverCommentsDeadline{XXXX}
 
\AtlasCoverAnalysisTeam{Filip Backman, Rebecca Carney, Karri Folan Di Petrillo, Melissa Franklin, Laura Jeanty, Lawrence Lee, Federico Meloni, Christian Ohm, Hideyuki Oide, Hidetoshi Otono, Simone Pagan Griso, Jennifer Kathryn Roloff, Abner Soffer, Emily Thompson, Alexander Naip Tuna}
 
\AtlasCoverEdBoardMember{Stephane~Willocq~(chair)}
\AtlasCoverEdBoardMember{Nicholas~Styles}
\AtlasCoverEdBoardMember{Philippe~Mermod}
 
\AtlasJournal{Physical Review D}
\AtlasJournalRef{Phys. Rev. D 102, (2020) 032006}
\AtlasDOI{10.1103/PhysRevD.102.032006}

\AtlasCoverEgroupEditors{atlas-SUSY-2018-33-editors@cern.ch}
 
\AtlasCoverEgroupEdBoard{atlas-SUSY-2018-33-editorial-board@cern.ch}
 
\hypersetup{pdftitle={ATLAS document},pdfauthor={The ATLAS Collaboration}}
 
\begin{document}
 
\maketitle

\clearpage
 
\section{Introduction}
 
The Standard Model of particle physics (SM) has successfully predicted the results of decades of laboratory experiments with impressive precision, but it suffers from several notable inadequacies. For example, the SM lacks explanations for the scale hierarchy of the interactions~\cite{Barbieri:1987fn,deCarlos:1993yy} and dark matter~\cite{Bertone:2004pz}, and it does not include a quantum description of gravity. However, despite ambitious search programs, the experiments at the Large Hadron Collider (LHC) have not yet reported any evidence of physics beyond the Standard Model (BSM).
 
A possible way for BSM signatures to evade the constraints from these searches is if the BSM particles produced in proton--proton ($pp$) collisions do not decay promptly but have lifetimes sufficiently long to yield decay lengths of the order of 1~mm or more. Such long-lived particles (LLPs) can generate a variety of unconventional detector signatures that are often completely free of irreducible backgrounds from SM processes. However, without dedicated reconstruction algorithms and analysis techniques that consider this possibility, a discovery of LLPs may be missed~\cite{Fairbairn:2006gg}. While there is a long history of searches for LLP signatures at colliders, such searches have attracted increased interest since the startup of the LHC~\cite{Lee:2018pag}.
This paper reports a search for decays of LLPs including a muon, using the full Run-2 dataset of the ATLAS experiment.
 
Extensions of the Standard Model involving supersymmetry (SUSY)~\cite{Golfand:1971iw,Volkov:1973ix,Wess:1974tw,Wess:1974jb,Ferrara:1974pu,Salam:1974ig} are appealing from a theoretical perspective, e.g.\ due to their potential to achieve gauge-coupling unification, provide an explanation for dark matter, and alleviate the naturalness problem~\cite{Sakai:1981gr,Dimopoulos:1981yj,Ibanez:1981yh,Dimopoulos:1981zb}. 
Scenarios with a SUSY partner of the top quark, the \emph{top squark} \stop, with a mass close to the weak scale are of particular interest. This is due to the large quantum corrections to the Higgs boson mass from top-quark loops that are at the center of the naturalness problem~\cite{Inoue:1982pi,Ellis:1983ed}.
 
The vast majority of searches for the \stop\ squark have assumed that $R$-parity is conserved. This quantity is defined as $R_p\equiv(-1)^{3(B-L)+2s}$ where $B$, $L$, and $s$ denote baryon number, lepton number, and spin, respectively~\cite{Barbier:2004ez}. Apart from small nonperturbative effects~\cite{tHooft:1976snw,Kuzmin:1985mm}, $B$ and $L$ are conserved in the SM. It is often assumed that their conservation will translate to the SUSY sector to automatically avoid low-energy constraints on $B$- and $L$-violation. However, these conserved quantities in the SM are due to accidental symmetries, not fundamental symmetries. In the minimal supersymmetric extension to the SM (MSSM)~\cite{Fayet:1976et,Fayet:1977yc}, couplings that violate baryon-number and lepton-number conservation naturally occur at tree level. The couplings responsible for these violations are collectively called $R$-parity violating (RPV) couplings. The RPV terms of the MSSM superpotential are given by
 
\begin{equation*}
{\mathcal W}_{\textrm{RPV}}=\mu_i l_i h_u + \lam_{ijk} l_i l_j\bar e_k +\lamp_{ijk}l_iq_j\bar d_k +\lampp_{ijk}\bar u_i\bar d_j\bar d_k,
\end{equation*}
 
\noindent where $\mu_i$, $\lambda_{ijk}$, $\lambda^{\prime}_{ijk}$, and $\lambda^{\prime\prime}_{ijk}$ are the RPV couplings, $l$ and $e$ represent the lepton and charged-lepton supermultiplets, $h_u$ represents the up-type Higgs supermultiplets, and $q$, $u$, and $d$ represent the quark, up-type quark, and down-type quark supermultiplets, respectively~\cite{Barbier:2004ez}.  The symbols $i$, $j$, and $k$ are generation indices. Nonzero RPV couplings can, for example, result in an unstable lightest SUSY particle (LSP) rendering constraints from many SUSY searches invalid. In such models, the LSP does not serve as a dark-matter candidate as it often does in $R_p$-conserving models~\cite{Goldberg:1983nd,Ellis:1983ew}.
 
\begin{figure}[!bt]
\begin{center}
\includegraphics[width=0.4\textwidth]{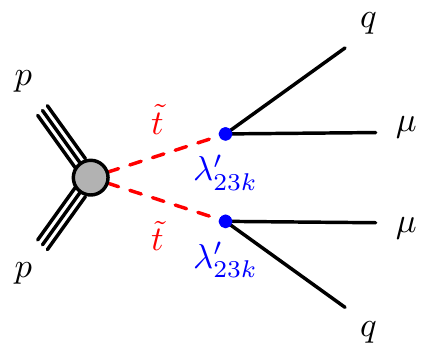}
\caption{\label{fig:signaldiagram} Diagram showing production of a top squark--anti-squark pair (both denoted by \stop), in which the top (anti-)squark decays into $\mu^{+}$ ($\mu^{-}$) and a down-type (anti-)quark of generation $k$. With sufficiently small values of the $R$-parity-violating coupling $\lambda^{\prime}_{23k}$, the lifetime of the $\tilde{t}$ becomes long enough to give rise to decays which are significantly displaced from their production point.}
\end{center}
\end{figure}

The search presented here targets nonzero values of the \lampijk\ coupling. All other RPV couplings are assumed to be exactly zero. Various low- and intermediate-energy constraints set upper limits on the size of these $\lampijk$ couplings, particularly for couplings involving light flavors. These include measurements of the elements of the Cabibbo–Kobayashi–Maskawa matrix, constraints on neutrinoless double-beta decay, and $B$-physics measurements.
Many of these constraints suggest that the value of any nonzero RPV coupling needs to be small.
In turn, this naturally leads to suppression of the decay processes and can give rise to long-lived SUSY particles. A nonzero $\lamp_{23k}$ coupling would allow a top squark to decay into a muon and a $k${\nth}-generation down-type quark, as shown in \Fig{\ref{fig:signaldiagram}}. The strongest indirect constraints on this coupling come from partial-width measurements of the $Z$ boson at LEP and exclude $\lamp_{23k} > 0.45$, assuming the existence of a squark with a mass of $100$~\GeV. A summary of experimental constraints on RPV SUSY is given in Ref.~\cite{Barbier:2004ez}.
 
In models with sufficiently small $\lambda^{\prime}_{23k}$ coupling values and where the \stop\ squark is the LSP, the suppression of the decay causes it to occur at discernible distances from the $pp$ interaction point where the \stop\ squark pair was produced. This would give rise to muons and high-mass vertices that are significantly displaced from the interaction point, yielding a distinctive detector signature in a collider experiment, with no irreducible backgrounds from SM processes. The search presented here is designed to be sensitive to this signature.
 
Other proposed BSM scenarios that could result in long-lived particle decays into at least one muon include long-lived lepto-quarks~\cite{Friberg:1997nn}, long-lived BSM particles appearing in decays of Higgs bosons~\cite{Strassler:2006ri,Cui:2014twa}, scenarios with right-handed neutrinos with Majorana masses below the electroweak scale~\cite{Asaka:2005pn} and RPV scenarios with a long-lived electroweakino LSP decaying through a virtual scalar muon and a $\lamp_{2jk}$ coupling~\cite{Allanach:2006st}.
 
Searches for a \stop\ squark decaying promptly via the \lampijk\ couplings have been performed by the ATLAS and CMS collaborations~\cite{Aaboud:2017opj,Sirunyan:2018ryt}. Exclusion limits on long-lived top squarks decaying into a muon and a hadronic jet have also been obtained by the CMS Collaboration, excluding \stop\ squark masses below $1.4$~\TeV\ for a mean proper lifetime of $0.1$~ns~\cite{Sirunyan:2017jdo,Khachatryan:2014mea,Sirunyan:2018vlw}. Related searches for displaced lepton production in association with displaced hadronic activity have been performed by the ATLAS Collaboration~\cite{Aad:2011zb,Aad:2012zx}.
 
In the search presented in this paper, dedicated tracking and vertexing algorithms are employed to retain selection efficiency for such LLP signatures. Events are required to contain a reconstructed displaced vertex and a reconstructed muon with a large impact parameter. Unlike previous searches from the ATLAS Collaboration~\cite{Aad:2011zb,Aad:2012zx}, the muon is not required to be associated with the vertex, in order to be inclusive of various BSM decay topologies. Signal region criteria are designed to select no more than roughly one background event in the available dataset, while maximizing the expected signal yield. The uncertainty in any background prediction is dominated by the statistical component. The predictions of the background yields are entirely derived in data, with individual contributions estimated from dedicated control regions.
 
\section{ATLAS detector}
\label{sec:detector}
 
The ATLAS detector~\cite{PERF-2007-01} at the LHC is a multipurpose particle detector with a forward--backward symmetric cylindrical geometry and a near \(4\pi\) coverage in solid angle.\footnote{ATLAS uses a right-handed coordinate system with its origin at the nominal interaction point (IP) in the center of the detector and the \(z\)-axis along the beam pipe. The \(x\)-axis points from the IP to the center of the LHC ring, and the \(y\)-axis points upwards. Cylindrical coordinates \((r,\phi)\) are used in the transverse plane, \(\phi\) being the azimuthal angle around the \(z\)-axis. The pseudorapidity is defined in terms of the polar angle \(\theta\) as \(\eta = -\ln \tan(\theta/2)\). Angular distance is measured in units of \(\Delta R \equiv \sqrt{(\Delta\eta)^{2} + (\Delta\phi)^{2}}\).}
It consists of an inner detector (ID) surrounded by a thin superconducting solenoid, electromagnetic and hadronic calorimeters, and a muon spectrometer (MS) incorporating three large superconducting toroidal magnets.
 
The ID is immersed in a \SI{2}{\tesla} axial magnetic field and provides charged-particle tracking in the range $|\eta| < 2.5$. The high-granularity silicon pixel detector covers the vertex region and typically provides four measurements per track, the first hit being normally in the insertable $B$-layer~\cite{ATLAS-TDR-19,Abbott:2018ikt} located just outside the beam pipe at a radius of 33~mm from the proton beam line. Three more concentric pixel layers are situated at radii of 50.5~mm, 88.5~mm and 122.5~mm before the first active layer of the semiconductor tracker (SCT) at $r = 299$~mm. Based on silicon microstrip technology, the SCT typically provides eight separate measurements. These silicon detectors are complemented by the transition radiation tracker (TRT), a straw-tube drift chamber which enables radially extended track reconstruction up to \(|\eta| = 2.0\).
 
In the central pseudorapidity range, energy measurements are provided by a high-granularity lead/liquid-argon (LAr) sampling calorimeter for electromagnetic deposits, and a steel/scintillator-tile calorimeter for hadrons ($|\eta| < 1.475$ and $|\eta| < 1.7$, respectively). The endcap and forward regions are instrumented with LAr calorimeters for both the electromagnetic and hadronic energy measurements up to \(|\eta| = 4.9\). Together, these systems provide full coverage in the region \(|\eta| < 4.9\).
 
The MS surrounds the calorimeters and features three large air-core toroidal superconducting magnets with eight coils each. The integral of the magnetic field from the toroids ranges between \num{2} and \SI{6}{\tesla\metre} for particle trajectories across most of the detector. The MS includes a system of precision tracking chambers and fast detectors for triggering. Three layers of drift-tube chambers provide an accurate measurement of the muon track curvature in the region \(|\eta| < 2.0\), while two layers of drift-tube chambers and one layer of cathode-strip chambers extend this measurement to \(|\eta| < 2.7\). Resistive-plate chambers provide muon triggering capability for \(|\eta| < 1.05\) while thin-gap chambers are used in the region \(1.05 < |\eta| < 2.4\).
 
Interesting events are selected to be recorded by the first-level trigger system implemented in custom hardware, followed by selections made by algorithms implemented in software in the high-level trigger~\cite{TRIG-2016-01}. The first-level trigger makes decisions at the 40~MHz bunch crossing rate to keep the event selection rate below \SI{100}{\kHz}, which the high-level trigger further reduces in order to record events to disk at about \SI{1}{\kilo\Hz}.
 
\section{Data samples and simulated events}
\label{sec:samples}
 
The analysis is performed on a set of $pp$ collision data at $\sqrt{s} = 13$~\TeV\ recorded during 2016--2018 which, after requiring good quality of the data~\cite{DAPR-2018-01}, corresponds to an integrated luminosity of $\intlumi$. The uncertainty in the combined 2015--2018 integrated luminosity is 1.7\% \cite{ATLAS-CONF-2019-021}, obtained using the LUCID-2 detector \cite{LUCID2} for the primary luminosity measurements. 
The data analyzed for this result were recorded using triggers requiring either a track in the MS with transverse momentum $\pt>60$~\GeV\ and $|\eta|<1.05$, or large missing transverse momentum $\met$ as measured in the calorimeters ($\MET > 100$~\GeV\ for a portion of the 2016 data, and $\MET > 110$~\GeV\ for the rest of the analyzed dataset). In the recorded events, there are approximately 35~$pp$ collisions in the same LHC bunch crossing, on average.
 
Samples of long-lived top squark--anti-squark pairs (denoted by $\stop\stop$) from Monte Carlo (MC) simulations are used as benchmarks to study expected signal efficiencies. Samples with 1~\TeV\ $\leq m(\stop) \leq$ 2~\TeV\ were generated in steps of $100$~\GeV, for mean proper lifetimes $\tau(\stop)$ of 0.01, 0.1 and 1~ns for each mass value. All other SUSY-particle contributions are assumed to be decoupled. The matrix element calculation for the \stop\ squark pair-production was performed to leading-order precision with \MGMCatNLOV{2.6.1}~\cite{Alwall:2014hca} with up to two additional outgoing partons, while the modeling of parton showers, hadronization and the underlying event was performed by \PYTHIAV{8.230}~\cite{Sjostrand:2014zea} with the A14 set of tuned parameters~\cite{ATL-PHYS-PUB-2014-021}. Parton distribution functions (PDF) from the NNPDF23LO~\cite{Ball:2012cx} set were used. In the signal models considered in this paper, the \stop\ squark lifetime is larger than the hadronization timescale in quantum chromodynamics (QCD). Since the \stop\ squark carries color charge, it will undergo a hadronization process in \PYTHIA and form a composite color-singlet state with SM quarks, an \emph{$R$-hadron}. Dedicated routines for hadronization of heavy colored particles~\cite{Farrar:1978xj} were used to simulate the hadronization process. The top squarks primarily form meson-like states ($\stop \bar{q}$), but approximately 10\% of them form baryon-like states ($\stop qq$). Roughly half of the $R$-hadrons formed around the top squarks have nonzero electric charge, and due to the $\stop\stop$ production the two $R$-hadrons cannot have electric charges of the same sign.
 
The \rhadron\ from \PYTHIA is then propagated through a simulation of the ATLAS detector~\cite{SOFT-2010-01} implemented in \GEANT 4 v10.1.3~\cite{Agostinelli:2002hh} employing dedicated models of $R$-hadron interactions with the detector material~\cite{Kraan:2004tz,Mackeprang:2006gx,Mackeprang:2009ad}, which can alter the content of the light-quark system in the $R$-hadron, possibly changing its electric charge as it traverses the detector. At the position of the decay, the \rhadron\ is passed to an instance of \PYTHIA which simulates the decay into final-state particles using the parton shower model described above. The resulting decay products are then further propagated through the detector simulation, starting from the point where the \rhadron\ decays. Reference~\cite{ATL-PHYS-PUB-2019-019} contains more complete technical details of the treatment of $R$-hadrons in the simulations. The decay process $\stop\rightarrow \mu q$ is simulated, where $q$ represents light-flavor, down-type quarks, given by nonzero $\lamp_{231}$ and $\lamp_{232}$ couplings. While the simulated mean proper lifetimes of the \stop\ squark are $0.01$, $0.1$, and $1$~ns, in the interpretation of the results, additional lifetime values are evaluated by reweighting events from these samples.
 
Signal production cross sections are calculated to approximate next-to-next-to-leading order in the strong coupling constant, adding the resummation of soft gluon emission at next-to-next-to-leading-logarithm accuracy (approximate NNLO+NNLL)~\cite{Beenakker:2016lwe,Beenakker:2014sma,Beenakker:2013mva,Beenakker:2011sf,Beenakker:2009ha,Kulesza:2009kq,Kulesza:2008jb,Beenakker:1996ch}. The nominal cross sections and their uncertainties are derived using the PDF4LHC15\_mc PDF set, following the recommendations of Ref.~\cite{Butterworth:2015oua}, and range from $6.8\pm0.8$~fb at a $\stop$ mass of $1$~\TeV\ to $15\pm4$~ab at $2$~\TeV.
 
Samples of $Z \rightarrow \mumu$ events generated using \POWHEGBOXV{v1 r2856}~\cite{Nason:2004rx,Frixione:2007vw,Alioli:2010xd,Alioli:2008gx} and \PYTHIAV{8.186}~\cite{Sjostrand:2007gs}, and reconstructed with the same configuration as the signal samples (see \Sect{\ref{sec:reco}}), are used to determine scale factors accounting for small differences in muon identification efficiency between the data and the MC simulation.
 
All simulated event samples include the effects of pileup, i.e.\ multiple $pp$ interactions per bunch crossing as well as effects on the detector response due to activity from bunch crossings before or after the one containing the selected event. These effects are modeled by overlaying simulated inelastic $pp$ events generated with \PYTHIAV{8.186}~\cite{Sjostrand:2007gs} using the NNPDF23LO set of PDFs~\cite{Ball:2012cx} and the A3 tune~\cite{ATL-PHYS-PUB-2016-017} over the original hard-scattering event.
 
\section{Event reconstruction}
\label{sec:reco}
 
The standard ATLAS track reconstruction is optimized for the trajectories of charged particles originating from the interaction region, or from decays of short-lived particles such as $b$-hadrons. To reduce the computational complexity, constraints are therefore placed on the transverse and longitudinal impact parameters of track candidates relative to the interaction region ($|d_0|<10$~mm and $|z_0|<250$~mm, respectively), and their hit multiplicities, to select particles emerging from the initial $pp$ collision. The primary vertex (PV) is defined as the collision vertex with the largest $\sum p_T^2$ of its associated tracks~\cite{ATL-PHYS-PUB-2015-026}.
 
In order to reconstruct tracks from the decays of long-lived BSM particles, which typically have impact parameters that fall outside the standard constraints, a dedicated track reconstruction is performed using \emph{large-radius tracking} (LRT)~\cite{ATL-PHYS-PUB-2017-014}. The LRT uses hits not already associated with tracks reconstructed by the standard tracking algorithm in order to improve the efficiency for tracks not originating from the interaction region. Requirements on the track impact parameters are relaxed to $|d_0| < 300$~mm and $|z_0| < 1500$~mm. Requirements on the number of hits which are allowed to be shared by multiple tracks are also relaxed. Tracks from the standard processing and the LRT processing are treated as a single collection in all subsequent steps of reconstruction, such as in the reconstruction of muons used in this search. Apart from the expanded input ID track collection, muons are reconstructed using the algorithms described in Ref.~\cite{PERF-2015-10}.
 
A dedicated secondary-vertex reconstruction algorithm for LLP decays is employed to reconstruct displaced vertices (DVs). As input to this vertex reconstruction, only tracks with transverse momentum $\pT > 1$~\GeV\ and $|d_0| > 2$~mm are considered. The $d_0$ requirement ensures that most tracks from $pp$ collisions and those produced in decays of short-lived, lower-mass SM particles such as $b$-hadrons, are ignored in DV reconstruction. In order to avoid contamination of the vertex reconstruction with fake tracks, which are reconstructed from spurious combinations of hits, the input-track selection includes several additional requirements. Tracks are required to have at least six~SCT hits or at least one pixel hit. A track is rejected if it has fewer than two pixel hits and also completely lacks TRT hits. Due to a larger relative contribution of fake tracks at low momentum, tracks with $\pT<25$~\GeV\ are subject to the additional requirements of having at least seven~SCT hits, as well as having at least 20~TRT hits if within $|\eta|<1.6$. Since only tracks with $|d_0| > 2$~mm are considered, a track from a charged LLP originating at the PV will not be included in the vertex reconstructed from its decay products.
 
The DV reconstruction begins by finding all possible two-track seed vertices from all possible pairs of selected tracks. The seed vertices are required to have a fit with $\chi^2$ per degree of freedom less than $5$. Several additional requirements are then applied to ensure the seed vertex tracks are compatible with the decay of a particle originating from the PV. Firstly, both tracks of the seed vertex are required not to have hits in pixel layers at radii smaller than the position of the seed vertex, and are required to have a hit in the first pixel or SCT layer at larger radius, if expected. If a seed vertex is inside or within several millimeters of a tracker barrel layer, hits in that particular layer are neither forbidden nor required. This avoids penalizing cases where it is unclear which side of the layer the vertex is on. Secondly, the inner product of the vector sum of the track momenta and the vector pointing from the PV to the seed vertex is required to be positive, ensuring that the seed vertex is pointing away from the PV.
 
The collection of all possible two-track seed vertices is then iteratively merged to form $n$-track vertices. Ambiguities due to tracks being compatible with more than one vertex are resolved by comparing their $\chi^2$ values in the vertex fits, or by merging the vertices if the distance between their estimated positions is not significant enough. To recover losses due to the tight requirements on the tracks used in the seed-vertex finding, tracks compatible with the reconstructed vertices are also attached even if they have associated hits in lower-radius layers. Extra requirements are made on the $d_0$ of the tracks before they are attached to the DV in order to suppress fake tracks from the reconstruction. More details about the secondary-vertexing algorithm can be found in Ref.~\cite{ATL-PHYS-PUB-2019-013}.
 
In the following, the number of associated tracks for a given DV is denoted by $\nTracks$, and its visible invariant mass calculated from the four-momenta of the associated tracks, assuming each track was produced by a charged pion, is denoted by $\mDV$.
 
The missing transverse momentum of an event is usually defined as the negative vector sum of the transverse momenta of all identified physics objects (electrons, photons, muons, jets) calibrated individually. In this analysis, an alternative definition using only topological clusters of energy deposits in the calorimeters is used instead~\cite{PERF-2014-07}. The clusters are locally calibrated to the hadronic scale, but no object-level information is used to calibrate the clusters. In the following, this is referred to as \emph{cluster-based} $\met$. This definition is used because of its similarity to the \met\ definition that is used in the trigger system, which is based solely on energy deposits in the calorimeter. The benchmark signal considered in this analysis does not include any genuine missing transverse momentum, but since high-\pt\ muons are expected to deposit only a small fraction of their energy in the calorimeters through ionization, most of their momentum is unaccounted for, making the \met\ trigger efficient for this signature.
 
\section{Event selection}
\label{sec:eventselection}
 
As the LRT processing is computationally expensive, it cannot be run on all events recorded with the ATLAS detector. Instead, during the prompt processing of the data, a tight selection filters out a subset of the events into a special raw-data stream. As the low-level hit information is ordinarily not stored for analysis, this special raw-data stream is later processed using the dedicated reconstruction configuration described in Section~\ref{sec:reco}. This filtering is therefore the first step in this analysis, defining the start of two mutually exclusive trigger-based event selections:
 
\begin{itemize}
\item \textbf{$\met$-triggered sample:} Events recorded with the $\met$ trigger were required to have cluster-based $\met > 180$~\GeV, in order to ensure that events are close to the plateau of the turn-on curve of the $\met$ trigger. The $\met$ trigger threshold varied in the range $100$--$110$~\GeV\ over the period in which the data were recorded.
\item \textbf{Muon-triggered sample}: Events recorded with the muon trigger, requiring a track in the MS with $|\eta| < 1.05$ and no explicit requirement of an ID track, are required to have at least one reconstructed muon with $\pT > 60$~\GeV. Reconstructed muons are not required to be matched to a muon that fired the trigger in the filtering step, although a trigger matching requirement is made after events are processed with special reconstruction. For events recorded during 2016 data-taking, MS tracks that had a well-matched ID track (without LRT) were required to have $|d_0| > 1.5$~mm. To make this filter more inclusive, for the events recorded during 2017--2018 data-taking, no requirement on the muon $d_0$ was imposed. The cluster-based $\met$ is required to be less than $180$~\GeV\ in order to make the two samples mutually exclusive.
\end{itemize}
 
Events that survive the filtering step detailed above are reconstructed with the standard ATLAS reconstruction algorithms and the special reconstruction described in \Sect{\ref{sec:reco}}. Events are then subject to additional selection criteria. Muon candidates are required to be reconstructed in both the ID (including LRT tracks) and the MS with $\pT>25$~\GeV\ and $|\eta|<2.5$. In order to reject muons that originate from SM particles, the $|d_0|$ of the muon with respect to the PV is required to be larger than 2~mm. Additionally, muon $|d_0|$ and $|z_0|$ are required to be less than $300$~mm and $500$~mm, respectively. These requirements define the \emph{muon preselection}.
 
Additional selection criteria are applied to preselected muons to reject major sources of background. The background sources identified for large-$d_0$ muons are cosmic-ray muons, reconstruction algorithm fakes, and muons from in-flight decays of SM hadrons, especially those containing heavy-flavor quarks.
 
To reduce the contribution from cosmic rays, events which have activity in the MS on the side opposite to the muon are rejected. The spatial positions of track segments reconstructed in individual muon stations are compared with the expected positions given the momentum direction of the reconstructed muon, assuming infinite momentum. Muons with matching segments on the opposite side of the MS that satisfy $|\eta(\textrm{muon momentum})+\eta(\textrm{segment position})|<0.05$ and $|\Delta\phi(\textrm{muon momentum, segment position})-\pi|<0.22$ are rejected. The difference between the $\phi$ and $\eta$ requirements is due to the MS detector geometry and its superior resolution in segment $\eta$. Track segments are used for this veto instead of fully reconstructed muon tracks, which are more likely to fail to be reconstructed as muons not pointing back to the interaction region. Angular corrections are applied to account for the longitudinal displacement of the muon. Muons are also rejected if they are opposite in $\eta$ and in $\phi$ to a region of the MS without detector coverage. This requirement is designed to ensure that if the muon were to pass through the detector like a cosmic-ray muon, it would pass through an instrumented region of the MS and can produce a track segment on the opposite side of the muon spectrometer. This requirement removes 94\% of the residual cosmic muon contribution and around 1\% of the contribution from typical signal samples. Together, these requirements constitute the \emph{cosmic-muon veto}.
 
Accidental reconstruction of fake displaced muons, from spurious combinations of hits, can also occur. Such fake muons, however, tend to have poor quality of fit and fewer hits on the track. To reject these fake muon candidates, they are required to be constructed from segments in at least three MS stations and have a quality of fit $\chi^2/N_{\textrm{DoF}}<8$. The latter requirement is kept loose to avoid MS alignment mismodeling effects. These requirements constitute the \emph{fake-muon veto}.
 
Muons produced in semileptonic decays of short-lived SM hadrons can also contribute to the background processes for this search. These are primarily decays of hadrons containing heavy-flavor quarks. However, such muons are most often produced with nearby energy depositions from hadronic activity. As a result, such processes can be rejected by requiring that the muons are isolated from nearby ID tracks and calorimeter energy deposits. The sum of \pt\ from tracks consistent with originating at the PV in a cone of varying size around the muon is required to be no more than $6\%$ of the muon \pt. The cone has a maximum size of $\Delta R=0.3$ for muons with $\pt<33.3$~\GeV, and a cone size of $10$~\GeV/$\pt$ is used for muons with higher $\pt$. Additionally, the sum of calorimeter cluster \pt in a cone of size $\Delta R=0.2$ around the muon is required to be no more than $6\%$ of the muon \pt. These two requirements define the \emph{heavy-flavor veto}.
 
Together, satisfying the cosmic-muon, fake-muon, and heavy-flavor veto criteria constitutes passing the \emph{full muon selection}.
 
The DVs are required to be reconstructed in a cylindrical volume with radius $\rDV<300$~mm and longitudinal extent $|\zDV|<300$~mm, have a fit $\chi^2/N_{\textrm{DoF}}<5$, and have a transverse distance from all reconstructed collision vertices greater than $4$~mm. To remain inclusive of other possible signals, there is no explicit requirement that a reconstructed muon be included in the vertex.
 
\begin{figure}[!htb]
\begin{center}
\includegraphics[width=0.45\textwidth]{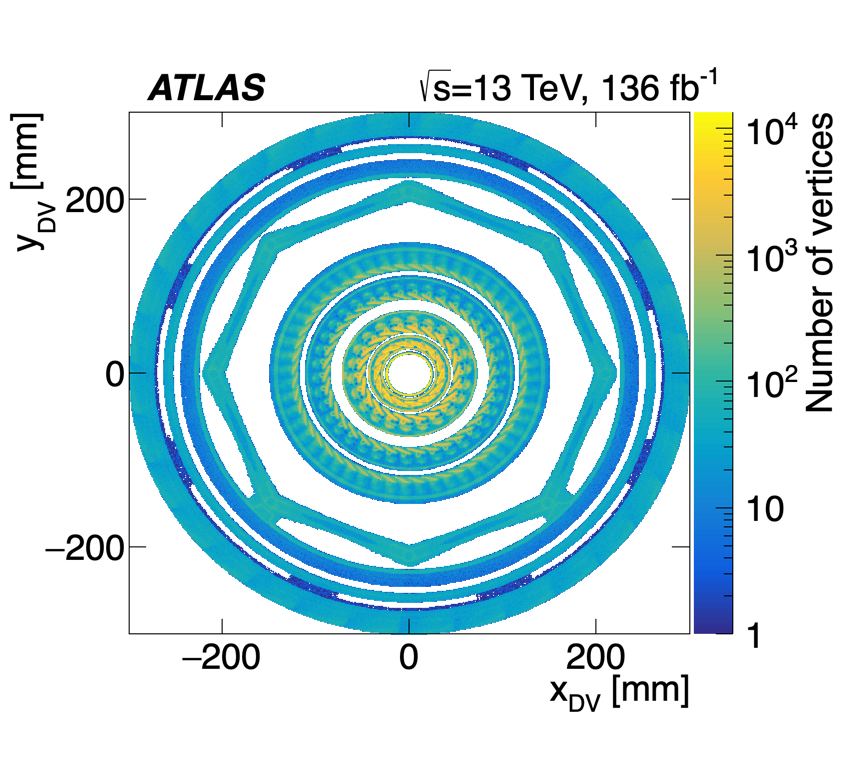}
\includegraphics[width=0.45\textwidth]{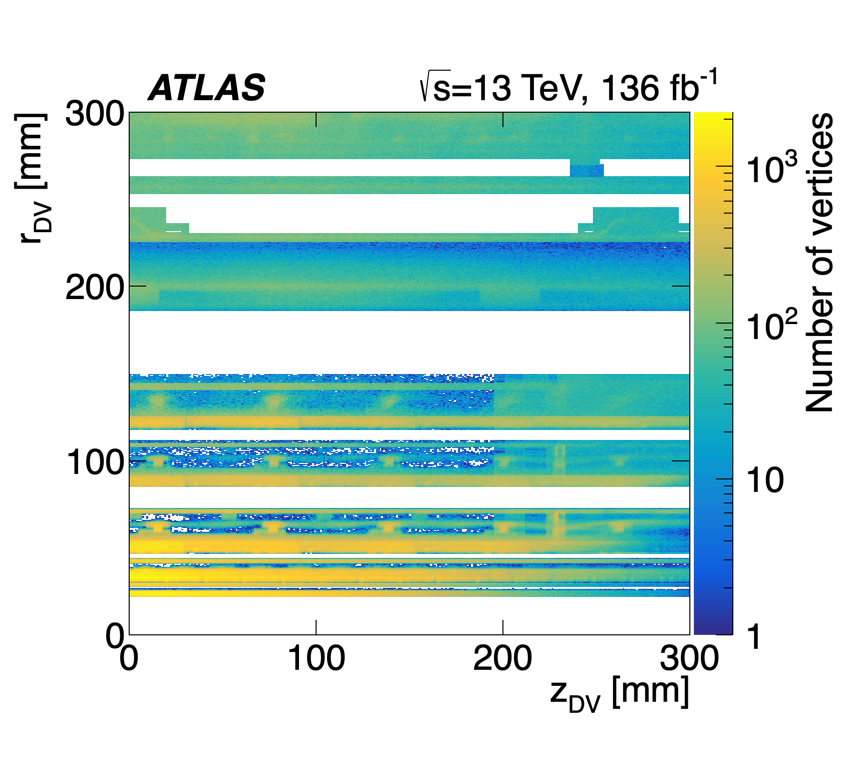}
\caption{The positions in the $x$--$y$ plane (left) and $r$--$z$ plane (right) of reconstructed displaced vertices that fail the material veto in the analyzed dataset. The latter figure is restricted to positive $z$ values for presentation.}
\label{fig:materialmap}
\end{center}
\end{figure}

To reject vertices arising from interactions with dense detector material, DVs are further required not to have a position consistent with sensitive elements of the detector, its support structures or its services. This veto is imposed via a three-dimensional map of detector material that is constructed from measured positions of low-mass vertices in an inclusive sample and the known positions of detector elements. This veto removes 42\% of the fiducial volume.
The positions of reconstructed vertices in the dataset that fail this material veto are shown in \Fig{\ref{fig:materialmap}}.
These requirements define the \emph{DV preselection}. DVs pass the \emph{full DV selection} if, in addition, they satisfy $\nTracks \geq 3$ and $\mDV > 20$~\GeV, requirements designed to reduce the expected number of background events to around one. For the signal scenarios considered in this paper, roughly half of the DVs satisfying the preselection criteria contain a track from a reconstructed muon, while roughly 80--90\% of DVs passing the full DV selection contain such a track.
 
\Tab{\ref{tab:selectionrequirements}} lists the preselection and full selection criteria for muons and DVs.
 
\begin{table}[!htb]
\begin{center}
\caption{Selection requirements for muons and displaced vertices.}
\begin{tabular}{l|c|c}
\noalign{\smallskip}\hline\hline\noalign{\smallskip}
\textbf{Selection level} & \textbf{Muon selection} & \textbf{Displaced vertex selection} \\
\hline
\rule{0pt}{2ex}Preselection 	& $\pT > 25$~\GeV, $|\eta| < 2.5$,	& $\rDV < 300$~mm, $|\zDV| < 300$~mm,\\
& 2~mm $< |d_0| <$~300~mm,		& $\min(|\vec{r}_{\mathrm{DV}} - \vec{r}_{\mathrm{PV}}|) > 4$~mm, $\chi^2/N_{\mathrm{DoF}} < 5$, \\
& $|z_0| <$~500~mm							& Pass material map veto \\
\noalign{\smallskip}\hline
\rule{0pt}{2ex}Full selection 	& Pass cosmic-muon, fake-muon,	& $\nTracks \geq 3$, \\
& and heavy-flavor vetoes			& $\mDV > 20$~\GeV \\
\noalign{\smallskip}\hline\hline\noalign{\smallskip}
\end{tabular}
\label{tab:selectionrequirements}
\end{center}
\end{table}
 
\begin{figure}[!htb]
\begin{center}
\includegraphics[width=0.45\textwidth]{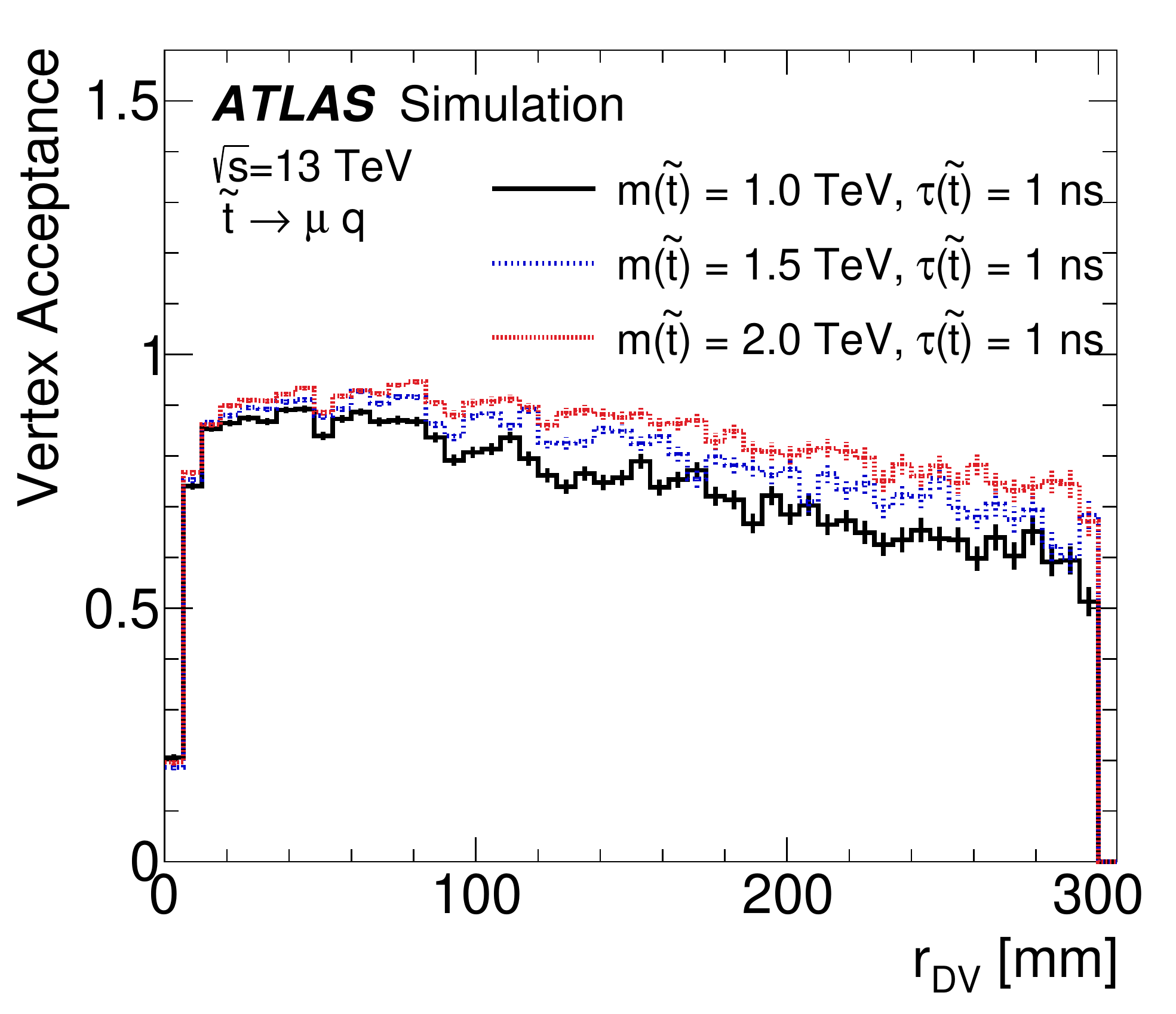}
\includegraphics[width=0.45\textwidth]{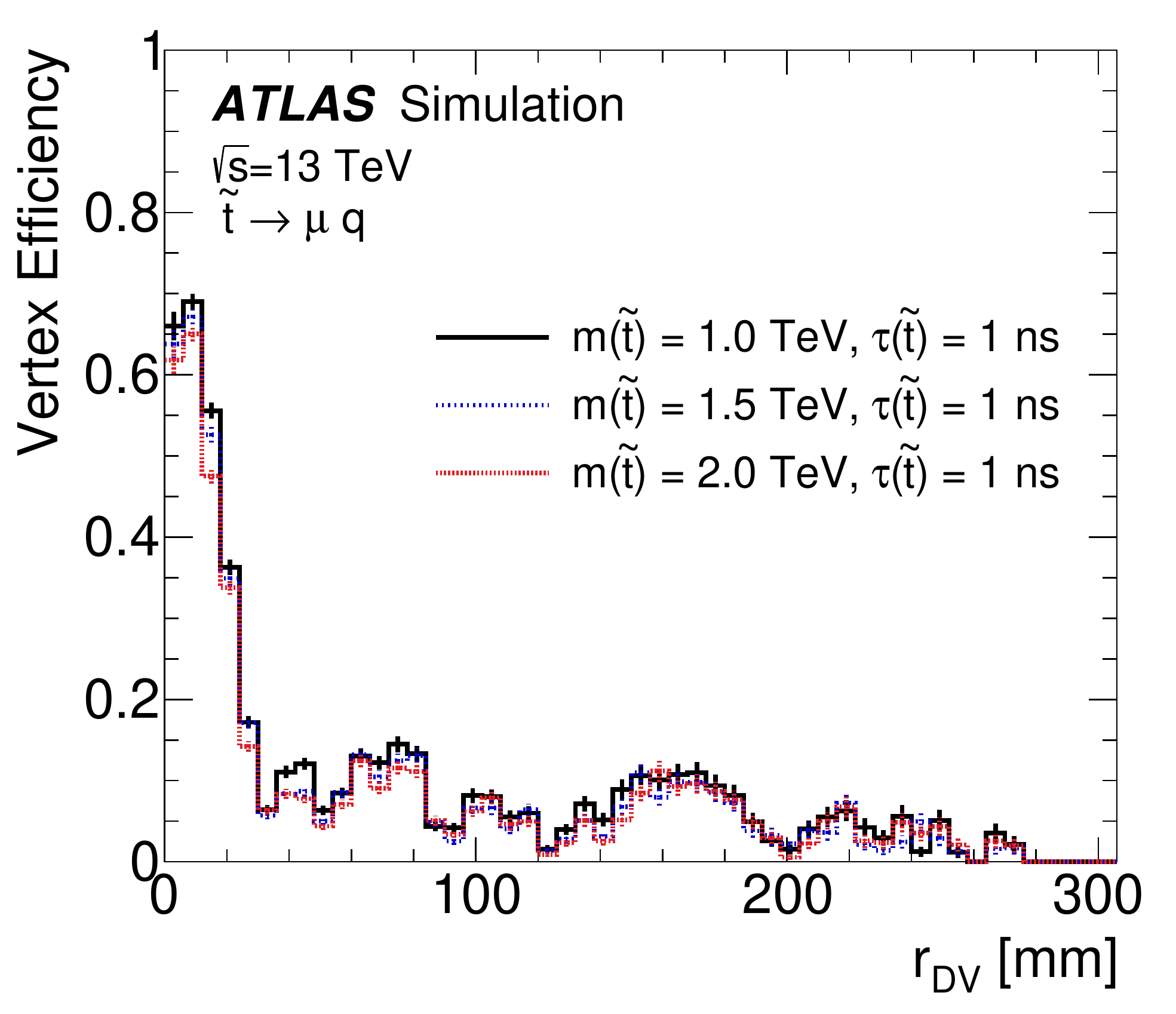}
\caption{Vertex selection acceptance (left) and efficiency (right) are shown for the $\tilde{t}$ \rhadron\ benchmark model as a function of the true transverse decay position \rDV.
}
\label{fig:vertexAcceptanceEfficiency}
\end{center}
\end{figure}

Figure~\ref{fig:vertexAcceptanceEfficiency} shows the vertex acceptance and efficiency for three benchmark signal scenarios with various mass hypotheses and $\tau(\stop)=1$~ns. The vertex acceptance is defined by requiring a $\tilde{t}$ \rhadron\ decay inside the fiducial volume considered by the analysis, $\rDV < 300$~mm, $|\zDV| < 300$~mm. \rhadron\ decays are further required to decay with a transverse position of $\rDV > 4$~mm, in order to emulate the requirement that displaced vertices be at least $4$~mm away from any $pp$ collision vertex in the event, in the transverse plane. Stable, electrically charged decay products of the \rhadron\ with $\pT>1$~\GeV\ and $|d_0|>2$~mm are used to compute the visible invariant mass and number of visible charged particles leaving the vertex. These requirements define the tracking acceptance. This visible invariant mass of the vertex is required to be greater than $20$~GeV, and the vertex must comprise at least three charged particles in the tracking acceptance. The vertexing efficiency is computed with respect to the \rhadron\ acceptance. In order to satisfy the efficiency requirement, \rhadron\ decays must be matched to a reconstructed DV that satisfies the full displaced vertex selection, as described in Table~\ref{tab:selectionrequirements}. The structure in the efficiency distribution as a function of \rDV\ is largely due to the application of the detector material veto.
 
For decays that occur inside the fiducial volume, the acceptance varies between $95\%$ and $55\%$. The only exception is for decays which occur within millimeters of the primary $pp$ vertex. These decays are either not sufficiently displaced from the primary vertex, or do not produce a sufficient number of charged particles with large transverse impact parameters.
Higher acceptance is observed for larger $\tilde{t}$ masses. The vertex efficiency is highest for decays occurring at radii smaller than the beam-pipe radius. For decays which occur at larger radii, the efficiency is affected by the tracking efficiency and material veto. Similar efficiency is observed for different $\tilde{t}$ masses.
 
The \emph{event preselection} is applied by requiring that events have at least one preselected muon and a PV with at least two tracks and $|z_{\mathrm{PV}}| < 200$~mm. Events in the signal regions (SRs) are required to have at least one fully selected muon and at least one fully selected DV, as described above. Two orthogonal SRs are defined, based on the trigger used to record events. The \emph{Muon Trigger SR} uses the MS track trigger described above and requires cluster-based \met\ below $180$~\GeV\ and a muon with $\pt>62$~\GeV\ and $|\eta|<1.05$ to ensure efficient triggering. This muon is further required to spatially coincide with the trigger-level muon. The \emph{\met Trigger SR} uses the \met-based trigger described above and requires that the cluster-based \met\ be larger than $180$~\GeV. The event selection requirements are summarized in \Tab{\ref{tab:eventselectionrequirements}}.
 
\begin{table}[!htb]
\caption{Event selection requirements.}
\begin{center}
\begin{tabular}{l|c|c}
\noalign{\smallskip}\hline\hline\noalign{\smallskip}
\textbf{Selection level} & \textbf{\met Trigger SR} & \textbf{Muon Trigger SR} \\
\hline
\rule{0pt}{2ex}Preselection 	& Selected by \met trigger, &  Selected by muon trigger,\\
& Cluster-based \met > 180~\GeV,  &  Cluster-based \met < 180~\GeV, \\
& Preselected muon & Preselected muon, \\
& & Highest-$\pT$ muon matches a trigger muon \\
& & and has $\pt>62$~\GeV\ and $|\eta|<1.05$ \\
\noalign{\smallskip}\hline
\rule{0pt}{2ex}Full selection 	& \multicolumn{2}{c}{$\geq1$ full-selection muon, $\geq1$ full-selection DV}	\\
\noalign{\smallskip}\hline\hline\noalign{\smallskip}
\end{tabular}
\end{center}
\label{tab:eventselectionrequirements}
 
\end{table}

\section{Background estimation}
\label{sec:bg}
 
The backgrounds for displaced muons described in \Sect{\ref{sec:eventselection}} are largely removed by dedicated veto requirements. Sources of background for DVs include detector material interactions and randomly intersecting tracks, which are efficiently suppressed by vetoing vertices in regions known to have material and requiring $\nTracks \geq 3$ and $\mDV>20$~\GeV.
 
The background estimation used in this analysis relies on the fact that the variables used to reject displaced muons from background sources are uncorrelated with the variables used to reject displaced vertices from background. This is exploited in order to estimate backgrounds in the SR from data. Any residual correlation is measured and treated as a systematic uncertainty in the background prediction as described in \Sect{\ref{sec:uncertainties}}. The following subsections describe how the residual backgrounds from cosmic rays, algorithm fakes, and heavy-flavor decays in the SRs are determined.
 
For each of the sources of background muons, a dedicated control region (CR) is constructed as described in \Sect{\ref{sec:regiondefinitions}}. Transfer factors for each of these muon backgrounds are measured in regions with background-like DVs and are used to extrapolate from CRs to the SR, as discussed in \Sect{\ref{sec:transferfactor}}. These extrapolations are validated in a set of validation regions (VRs) with an orthogonal set of background-like DVs. The full estimation of the background yield is presented in \Sect{\ref{sec:backgroundpredictions}}.
 
\subsection{Region definitions}
\label{sec:regiondefinitions}
 
Preselected events are divided into three regions depending on the DV properties:
\begin{itemize}
\item \textbf{DV Control Region (DV CR)}: no DVs passing the preselection, comprising three subregions:
 
\begin{itemize}
\item No DVs with a good fit quality passing the material map veto,
\item Only two-track DVs failing the material map veto,
\item Only three-track DVs failing the material map veto;
\end{itemize}
 
\item \textbf{DV Validation Region (DV VR)}: preselected DVs not passing the full selection, comprising two subregions:
 
\begin{itemize}
\item Preselected two-track DVs,
\item Preselected DVs with $\mDV<20$~\GeV\ and $\nTracks>2$; and
\end{itemize}
 
\item \textbf{DV Signal Region (DV SR)}: DVs passing the full selection.
\end{itemize}
 
For each muon background, a control region (CR) enriched with events of this background is defined by inverting the dedicated veto. A transfer factor is then determined as the ratio of the number of events passing the veto to the number of events rejected by it. As the probability of passing or failing the muon veto does not depend on the DV properties of the event, this transfer factor can be measured in the DV CR and applied in the DV SR in order to estimate the background contribution.
 
The following muon selections are used to define an orthogonal slicing of the dataset:
\begin{itemize}
\item \textbf{Fake-muon CR}: full muon selection with the fake-muon veto inverted,
\item \textbf{Heavy-flavor CR}: full muon selection with the heavy-flavor veto inverted,
\item \textbf{Cosmic-muon CR}: full muon selection with the cosmic-muon veto inverted, and
\item \textbf{Muon SR}: full muon selection.
\end{itemize}
 
The final signal regions used in this search are the intersections of the DV SR requirements and the Muon SR requirements.
 
\subsection{Transfer factor determination}
\label{sec:transferfactor}
 
The DV~CR is used to determine the transfer factors for each of the three muon backgrounds. The transfer factor $f_i$ is measured as
 
\begin{equation*}
f_i = \frac{ N(\mathrm{Preselected~events;DV\in DV~CR;\mu~\textbf{passes}~veto}~i~\mathrm{and~passes}~d_0~\mathrm{selection}~i )}
{ N(\mathrm{Preselected~events;DV\in DV~CR;\mu~\textbf{fails}~veto}~i~\mathrm{and~passes}~d_0~\mathrm{selection}~i )},
\end{equation*}
 
where $i$ represents the fake-muon, heavy-flavor, and cosmic-muon backgrounds, and events in both the numerator and denominator are required to pass the other two vetoes that compose the full muon selection. Varying $d_0$ selections are applied for each background estimation and are described below.
 
The fake-muon and heavy-flavor transfer factors are determined in the \met-triggered sample, but because a larger number of events with cosmic-ray muons are expected to be found in the muon-triggered sample, the corresponding transfer factor is determined in those events. Additionally, to minimize the impact of muons from heavy-flavor decays on the determination of the transfer factors for fake-muon and cosmic-muon backgrounds, those are determined using preselected muons with $5$~mm~$<|d_0|<300$~mm. For similar reasons, the heavy-flavor transfer factor is measured using preselected muons with $1.5$~mm~$<|d_0|<3.0$~mm. The transfer factor dependence on these requirements on muon $d_0$ are accounted for in a dedicated systematic uncertainty as discussed in \Sect{\ref{sec:uncertainties}}.
 
Given these selections, both the numerator and denominator of the transfer factors are taken from regions pure in their respective backgrounds, using the inversion of dedicated vetos as well as event and muon properties characteristic of each background source. \Tab{\ref{tab:transfer_factors}} reports the transfer factors extracted from the data in the DV CR, along with their uncertainties (discussed in \Sect{\ref{sec:uncertainties}}).
 
\begin{table}[!ht]
\caption{Transfer factors used to calculate background estimates, extracted from control regions.}\begin{center}
\begin{tabular}{lc}
\noalign{\smallskip}\hline\hline\noalign{\smallskip}
\textbf{Background source ($i$)} & \textbf{Transfer factor ($f_i$)} \\
\hline
Cosmic-ray muons & $(4.0 \pm 0.2 \textrm{ (stat.)} \pm 0.5 \textrm{ (syst.)} )\times 10^{-3}$ \\
Fake muons & $ (1.0 \pm 0.3 \textrm{ (stat.)}\pm 0.5\textrm{ (syst.)} )\times 10^{-2}$ \\
Heavy-flavor muons & $( 9.1 \pm 1.6 \textrm{ (stat.)} \pm 4.0\textrm{ (syst.)} )\times 10^{-2}$ \\
\noalign{\smallskip}\hline\hline\noalign{\smallskip}
\end{tabular}
\end{center}
\label{tab:transfer_factors}
 
\end{table}
 
Although the fake-muon and heavy-flavor transfer factors are measured in the \met-triggered sample, they are also applied in the muon-triggered sample. The cosmic-ray transfer factor is similarly applied to the \met-triggered sample. In order to ensure that the extrapolation between trigger streams does not lead to a bias, a separate measurement of the heavy-flavor transfer factor is made in the muon-triggered sample and compared with the nominal transfer factor. Agreement within uncertainties is observed. It is impossible to repeat this check with cosmic-ray and fake-muon transfer factors, because they are negligible backgrounds in the \met-triggered and muon-triggered samples respectively. Furthermore, cosmic-ray muons are expected to exhibit the same characteristics regardless of the trigger used to collect the event.

\subsection{Background predictions}
\label{sec:backgroundpredictions}
 
Using the transfer factors determined from the data, the background in the final SRs of the analysis can be predicted for each source $i$ by multiplying the event yield observed in the corresponding muon CR, $N^{\mathrm{CR}}_i$, in events with at least one DV passing the full selection (DV SR) by $f_i$. The total expected background $N^{\mathrm{SR}}_{\mathrm{B}}$ in each SR of the analysis is then given by
 
\begin{eqnarray*}
N^{\mathrm{SR} }_{\mathrm{B}} = \sum_i{N^{\mathrm{CR}}_i \times f_i}.
\end{eqnarray*}
 
Similarly, the transfer factors are validated by comparing their predictions with the observed data in the DV VR to gain confidence that the method works. The VR is divided into two separate regions, one with two-track DVs and one with low-mass DVs with three tracks or more. The observed results of this validation are discussed in \Sect{\ref{sec:results}}.
 
Using the technique described above, the total predicted background yields are \metSRYield for the \met Trigger SR and \muonSRYield for the Muon Trigger SR. CR yields used in the background estimation are shown in \Tab{\ref{tab:bg_estimate}} along with predictions for the SRs. For the \met Trigger SR, no events are observed in the Cosmic-muon CR, so fewer than $0.01$ cosmic-ray muon events are expected in the SR at 68\% confidence level (CL). This background component is therefore considered negligible in the presence of the other background contributions.
 
\begin{table}[!ht]
\caption{Control region yields $N^{\mathrm{CR}}_i$ are shown for each major background for the \met\ Trigger region and the Muon Trigger region, along with the result of scaling each background component by the appropriate transfer factor $f_i$. The total background yields $N^{\mathrm{SR} }_{\mathrm{B}}$ expected in each SR are also shown. The uncertainties shown are statistical and systematic, respectively. For the \met\ Trigger region, the contribution from cosmic-ray muons is found to be negligible as described in the text.}
\begin{center}
\begin{tabular}{llrccc}
\noalign{\smallskip}\hline\hline\noalign{\smallskip}
\textbf{Selection} & \textbf{Background source} & \textbf{$N^{\mathrm{CR}}_i$} & \textbf{$N^{\mathrm{CR}}_i\times f_i$} & \textbf{$N^{\mathrm{SR} }_{\mathrm{B}}$} \\
\noalign{\smallskip}\hline\noalign{\smallskip}
& 	Cosmic-ray muons 	& 	$0$ 	& 	$-$ &  \rdelim\}{3}{9em}[~~$0.43\pm0.16\pm0.16$] \\
\met\ Trigger	& 	Fake muons 			& 	$16$ 	& 	$0.16\pm0.04\pm0.09$ &\\
& 	Heavy-flavor muons 	& 	$3$ 	& 	$0.27\pm0.16\pm0.13$ & \\
\hline\noalign{\smallskip}
& 	Cosmic-ray muons 	& 	$376$ 	& 	$1.50\pm0.08\pm0.23$ & \rdelim\}{3}{9em}[~~$1.88\pm0.20\pm0.28$]\\
Muon Trigger	& 	Fake muons 	& 	$1$ 	& 	$0.01\pm0.01\pm0.01$ & \\
& 	Heavy-flavor muons 	& 	$4$ 	& 	$0.36\pm0.18\pm0.17$ & \\
\noalign{\smallskip}\hline\hline\noalign{\smallskip}
\end{tabular}
\end{center}
\label{tab:bg_estimate}
\end{table}

\section{Uncertainties}
\label{sec:uncertainties}
 
Uncertainties affecting the estimated SR event yields for the various sources of background are assessed by using transfer factors extracted in subregions of the CRs probing different regions of DV properties, and taking the largest deviation from the nominal transfer factor measurement. The DV CR is divided into three subregions of events: those with no reconstructed DV and those with two-track DVs or three-track DVs that do not pass the material veto. Given the differing DV properties, the span of these transfer factors represents the overall uncertainty from residual correlations between DV and muon properties. Uncertainties are also assessed by measuring transfer factors with varying requirements on the muon $d_0$. For the fake-muon and cosmic-ray transfer factors, separate measurements are evaluated for muons with $5 < |d_0| < 100$ mm and $100 <|d_0| < 300$ mm. For the heavy-flavor transfer factor, transfer factors are evaluated for muons with $1.5 < |d_0| < 1.7$ mm and $1.7 <|d_0| < 3$ mm. These impact parameter selections were designed to minimize the statistical uncertainties of the alternative transfer factor measurements. The spread of these transfer factors is measured to account for residual dependence of the transfer factor on the measured muon properties. The uncertainties from these two variations are added in quadrature. The uncertainties for the final SR background predictions quoted in Section~\ref{sec:bg} are determined in this way.
 
A number of factors affect the uncertainty of the event yields predicted for signal scenarios, and this section outlines how their impact on the final event selection efficiencies $\seleff$ is determined.
 
The largest overall uncertainty is related to the dedicated LRT and secondary-vertex reconstruction. The radial distributions of secondary vertices from \Kshort decays are compared in data and \PYTHIA multijet simulation, after normalizing the two distributions by the number of vertices at low radii, where tracking efficiencies are well understood. The mismodeling of the reconstruction efficiency in MC simulation at large track impact parameter and large vertex radius is estimated in this comparison. The largest observed difference is applied as a conservative per-track efficiency uncertainty of $\pm10\%$ which is then applied to the tracks associated with displaced vertices in the signal samples. The effect on final event counts with a fully selected DV is determined to be 15\%.
 
Additional uncertainty due to imperfect trigger efficiency is determined by comparing the $\met$ trigger efficiency turn-on as function of cluster-based $\met$ in $Z\rightarrow\mumu$ events in data and MC simulation. An $\met$-dependent correction (between 0 and 10\%) is extracted and applied to the MC samples for events which have cluster-based $\met$ between $180$ and $220$~\GeV. The relative difference between the final SR event yields with and without this correction is taken as a systematic uncertainty, found to be $<0.2$\%.
The uncertainties in the scale and resolution for the cluster energies entering the $\met$ calculation are assessed by studying the size of these uncertainties for fully reconstructed jets~\cite{PERF-2016-04}. The same variations are also observed to have a small effect on the muon isolation efficiency. The combined effect is determined to be no more than 2.1\%, which is applied to all signal efficiencies.
 
For the efficiency of the MS-only muon trigger, the relative efficiency degradation as a function of $d_0$ with respect to prompt muons as determined in MC signal samples is taken as an uncertainty. The per-track reconstruction efficiency uncertainty extracted above from \Kshort decays is also applied for the ID part of the muon track reconstruction. The muon reconstruction and selection criteria efficiency are also determined in MC, and the degradation as a function of $d_0$ with respect to prompt muons is taken as an uncertainty. The effects of these uncertainties are assumed to be uncorrelated. Their combined impact on the signal event yields in the SR is 10--12\%. The uncertainty due to residual inefficiency in the MS-only muon trigger given prompt muon production is found to be $<0.2$\%.
In addition, since custom muon identification criteria are applied, differences between data and MC events are corrected for by comparing efficiencies for prompt muons in $Z\rightarrow\mumu$ events in data and MC samples processed with the special reconstruction used for the analysis. The statistical uncertainties of these measured scale factors are propagated to the final SR event yield uncertainties for signal, resulting in an impact of 0.9--4.0\% on the total selection efficiency. This uncertainty is added in quadrature with standard prompt muon uncertainties due to mismeasurement and result in additional uncertainties below 1\%.
 
To assess the sensitivity to effects due to imperfect modeling of the pileup interactions in the signal samples, a reweighting is applied to the MC signal samples to cover the uncertainty in the number of $pp$ interactions in the data. The largest relative difference in $\seleff$ is determined to be between 0.37\% and 2.2\%. Another uncertainty stems from the modeling of initial-state QCD radiation (ISR), which affects the kinematics of the $\stop\stop$ system and thus the cluster-based $\met$ and the muon $\pT$ distribution. To assess the sensitivity of $\seleff$ to this, a reweighting of the $\pT$ distribution of the $\stop\stop$ system in the \MGMCatNLO events is applied to simulate that of \textsc{Pythia}~6.427. The resulting impact on $\seleff$ is 3\%. Finally, the uncertainty of $1.7\%$ in the integrated luminosity of the dataset is included as that translates directly to an uncertainty in the expected yield of signal events in the SRs.
 
The impact of the systematic uncertainties affecting the signal yield in both signal regions is summarized in \Tab{\ref{tab:sigsystsummary}}.
 
\begin{table}[!ht]
\caption{Summary of the impact of systematic uncertainties on the predicted yield for the signal scenarios with pair-produced long-lived top squarks. These uncertainties apply to both signal region selections used in the search unless otherwise specified.}
\begin{center}
\begin{tabular}{lr}
\noalign{\smallskip}\hline\hline\noalign{\smallskip}
Source of uncertainty & Relative impact signal yield [\%] \\
\noalign{\smallskip}\hline\noalign{\smallskip}
Total & $18$--$20$ \\
\noalign{\smallskip}\hline\noalign{\smallskip}
Tracking and vertex reconstruction & $15$ \\ 
Displaced muon efficiency & $10$--$12$ \\ 
Prompt muon efficiency & $(0.01$--$0.7)~\oplus~(0.9$--$4.0)$\\
ISR modeling in MC simulation& $3$\\
Pileup modeling & $0.37$--$2.2$ \\
Hadronic energy scale and resolution (affecting $\met$) & $2.1$\\ 
Integrated luminosity of dataset & $1.7$\\
Trigger efficiency & $<0.2$\\
\noalign{\smallskip}\hline\hline\noalign{\smallskip}
\end{tabular}
\end{center}
\label{tab:sigsystsummary}
\end{table}

\section{Results}
\label{sec:results}
 
No events are observed in the \met Trigger SR where \metSRYield events are expected in the absence of signal, while a single event is observed in the Muon Trigger SR where \muonSRYield events are expected. The observed event yields are in good agreement with the background-only expectations. The single event observed in the Muon Trigger SR contains a selected muon with $\pt=103$~\GeV, and two selected DVs. One DV has $\nTracks=3$, $\mDV=23.5$~\GeV, and net electric charge of $+1e$, while the other has $\nTracks=3$, $\mDV=22.7$~\GeV, and a net electric charge of $-1e$.
 
\begin{figure}[!tb]
\begin{center}
\includegraphics[width=0.45\textwidth]{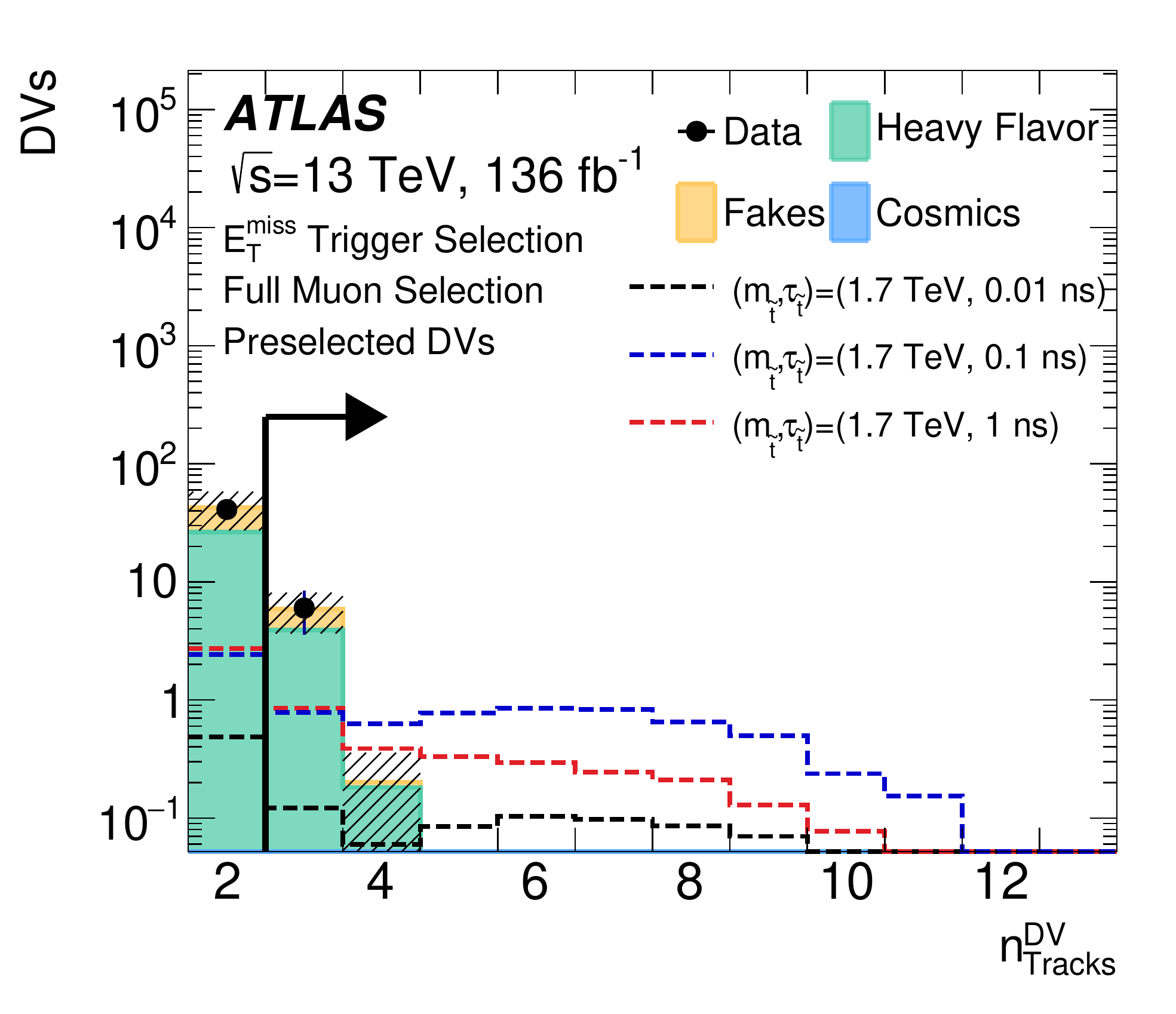}
\includegraphics[width=0.45\textwidth]{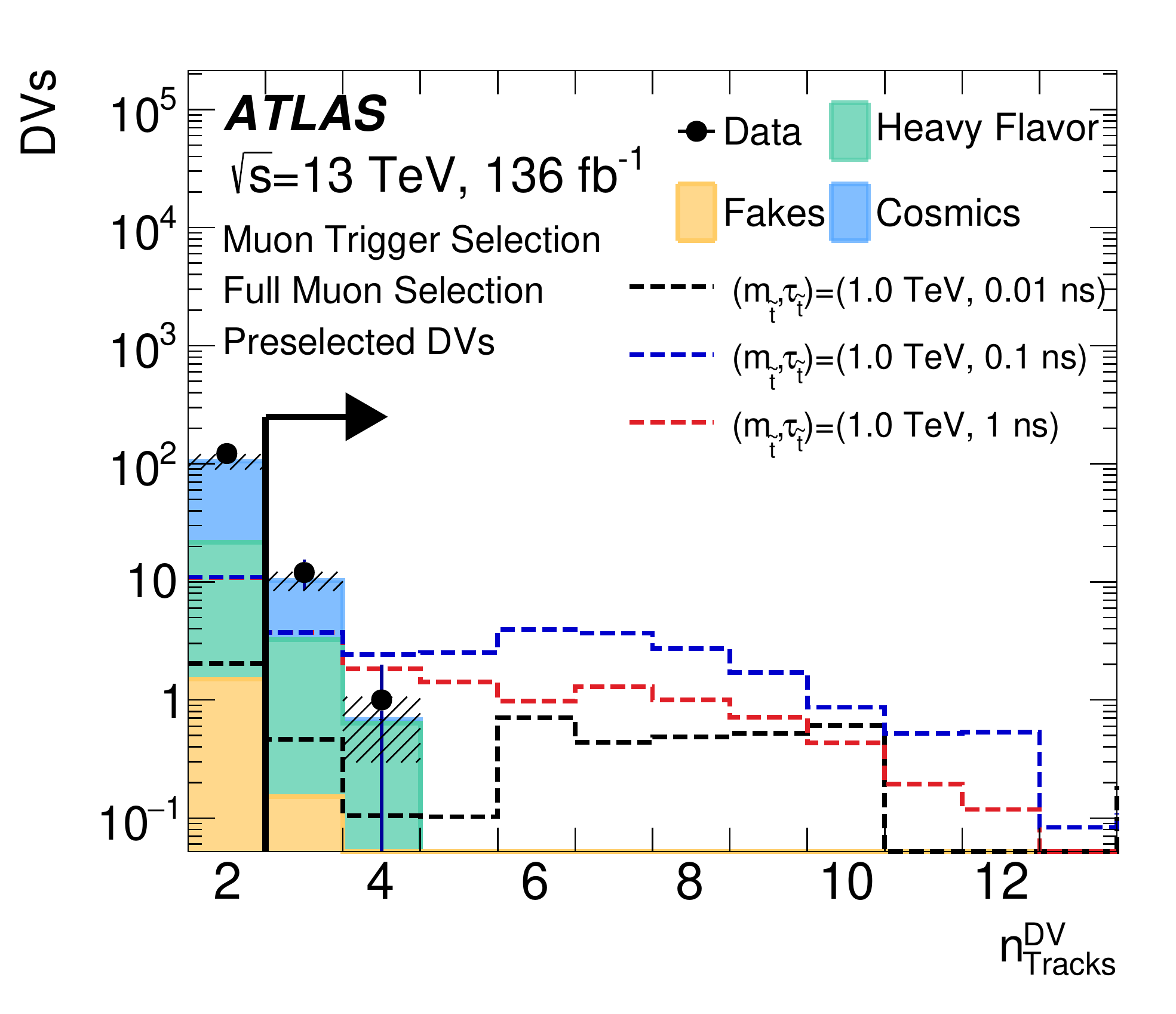}\\
\includegraphics[width=0.45\textwidth]{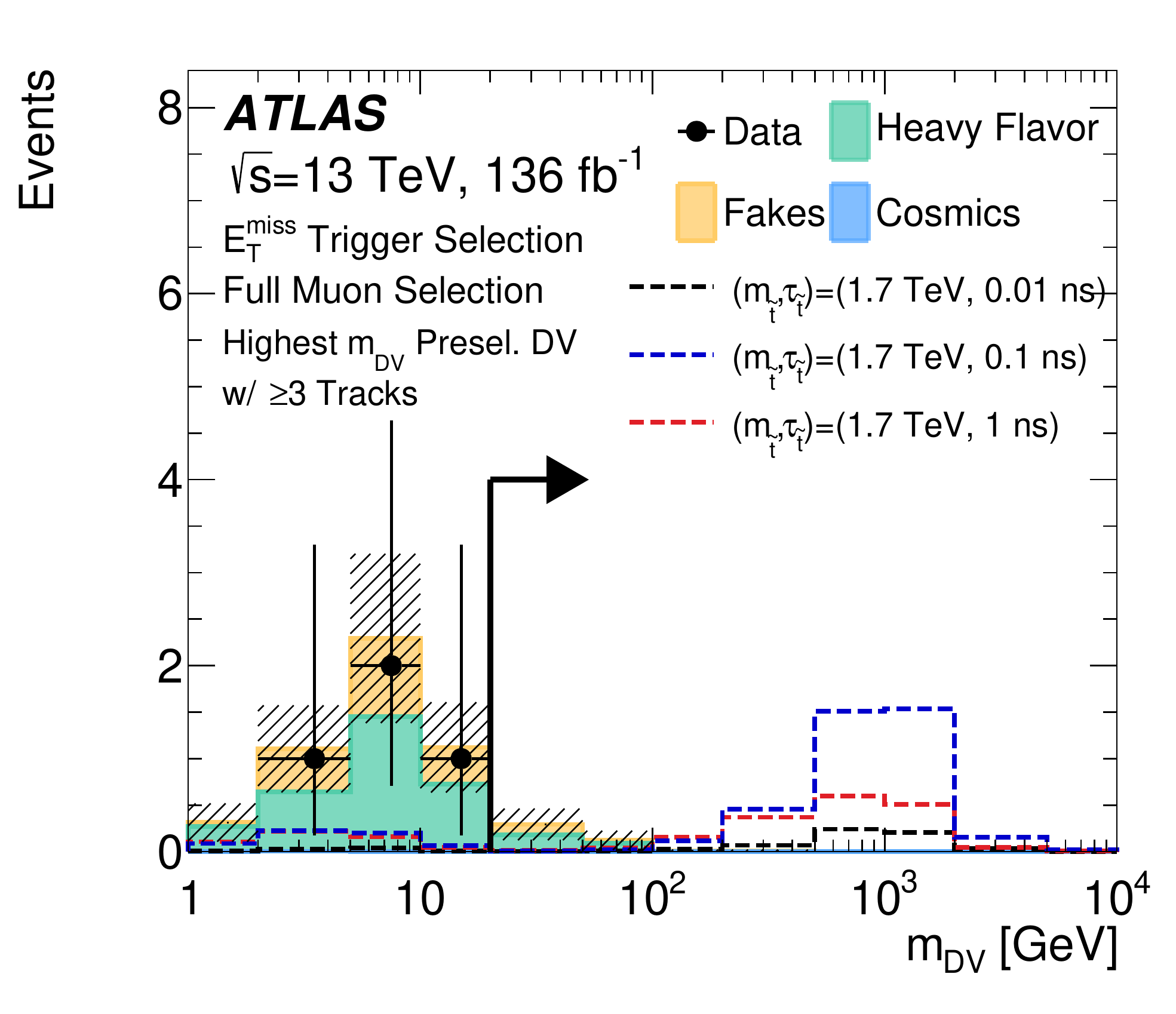}
\includegraphics[width=0.45\textwidth]{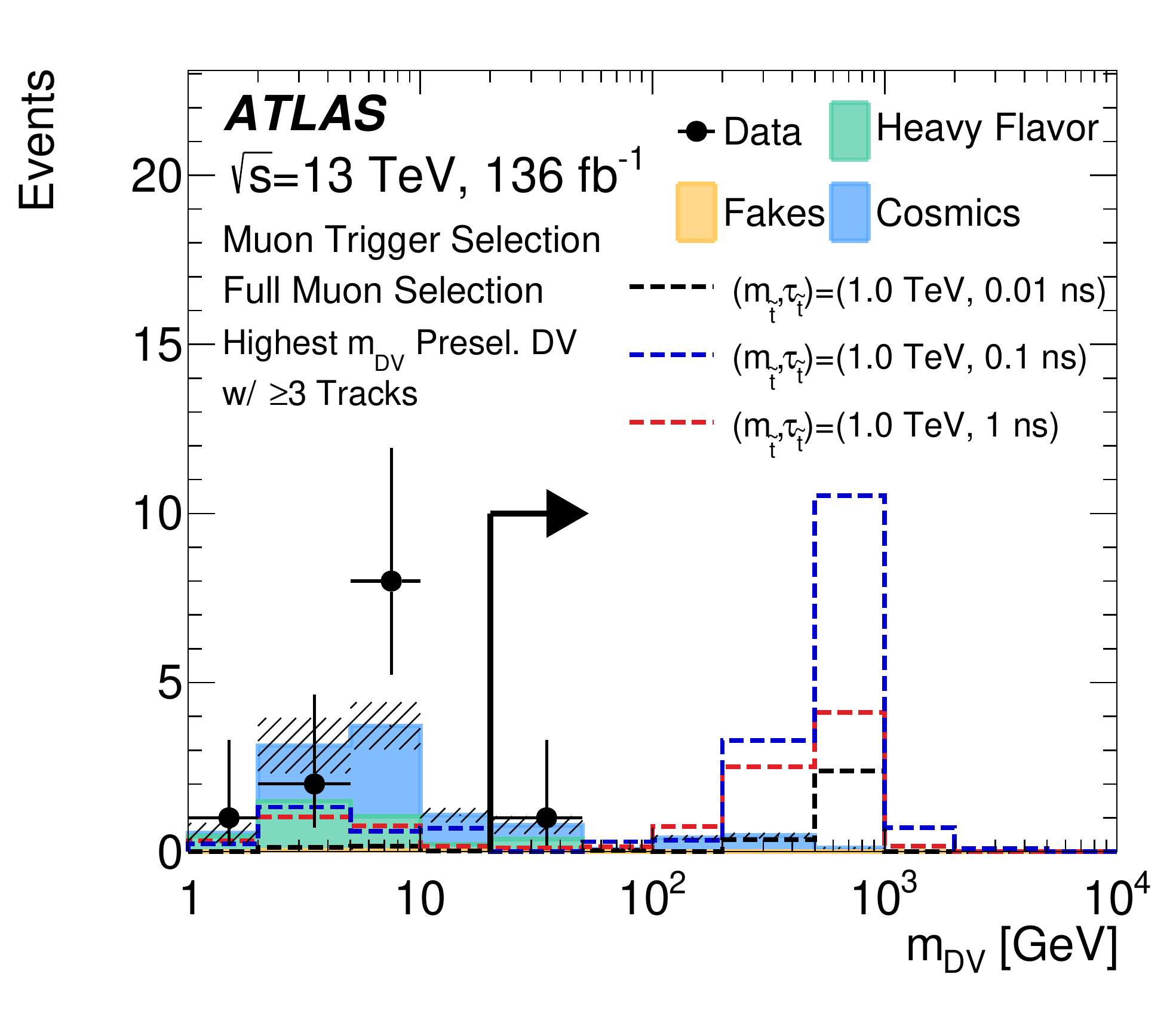}
\caption{Distributions of track multiplicity $\protect\nTracks$ (top) for preselected DVs, and invariant mass $\mDV$ (bottom) for the highest-mass preselected DV with at least three associated tracks. This is shown for events with at least one muon passing the full selection, in the $\met$-triggered (left) and muon-triggered (right) samples. Along with the data, shown with black markers, the stacked filled histograms represent the background estimates, and predictions for signal scenarios are overlaid with dashed lines. Signals with lifetimes of $0.01$~ns, $0.1$~ns, and $1$~ns are shown in black, blue, and red, respectively. For the $\met$-triggered (muon-triggered) selection, signal samples are shown with $m(\stop)=1.7$~TeV ($m(\stop)=1.0$~TeV). The errors include statistical and systematic uncertainties and are indicated by hatched bands. The DV full selection requirements, $\protect\nTracks \geq 3$ and $\protect\mDV > 20$~\GeV\ are visualized with a black arrow.}
\label{fig:stacked_backgrounds_dv_ntracks}
\end{center}
\end{figure}

\Fig{\ref{fig:stacked_backgrounds_dv_ntracks} } shows the distributions of the variables used for the final two DV-selection requirements in events with a muon passing the full muon selection. The track multiplicity $\nTracks$ is shown for all preselected displaced vertices. The invariant mass $\mDV$ is shown for the highest-mass preselected displaced vertex with at least three associated tracks. The individual background predictions are derived from muon CR distributions scaled by the transfer factors described in \Sect{\ref{sec:bg}}.
 
For all four distributions, there is good agreement between the number of vertices observed in data and the background prediction. The DV full selection requirements, $\nTracks \geq 3$ and $\mDV>20$~\GeV\ are visualized with a black arrow. For the \mDV\ distributions, inverting the requirement denoted by the black arrow gives the low-mass DV VR, showing good agreement between the expected background normalization and the observed number of events.

\FloatBarrier
 
\begin{figure}[!tb]
\begin{center}
\includegraphics[width=0.45\textwidth]{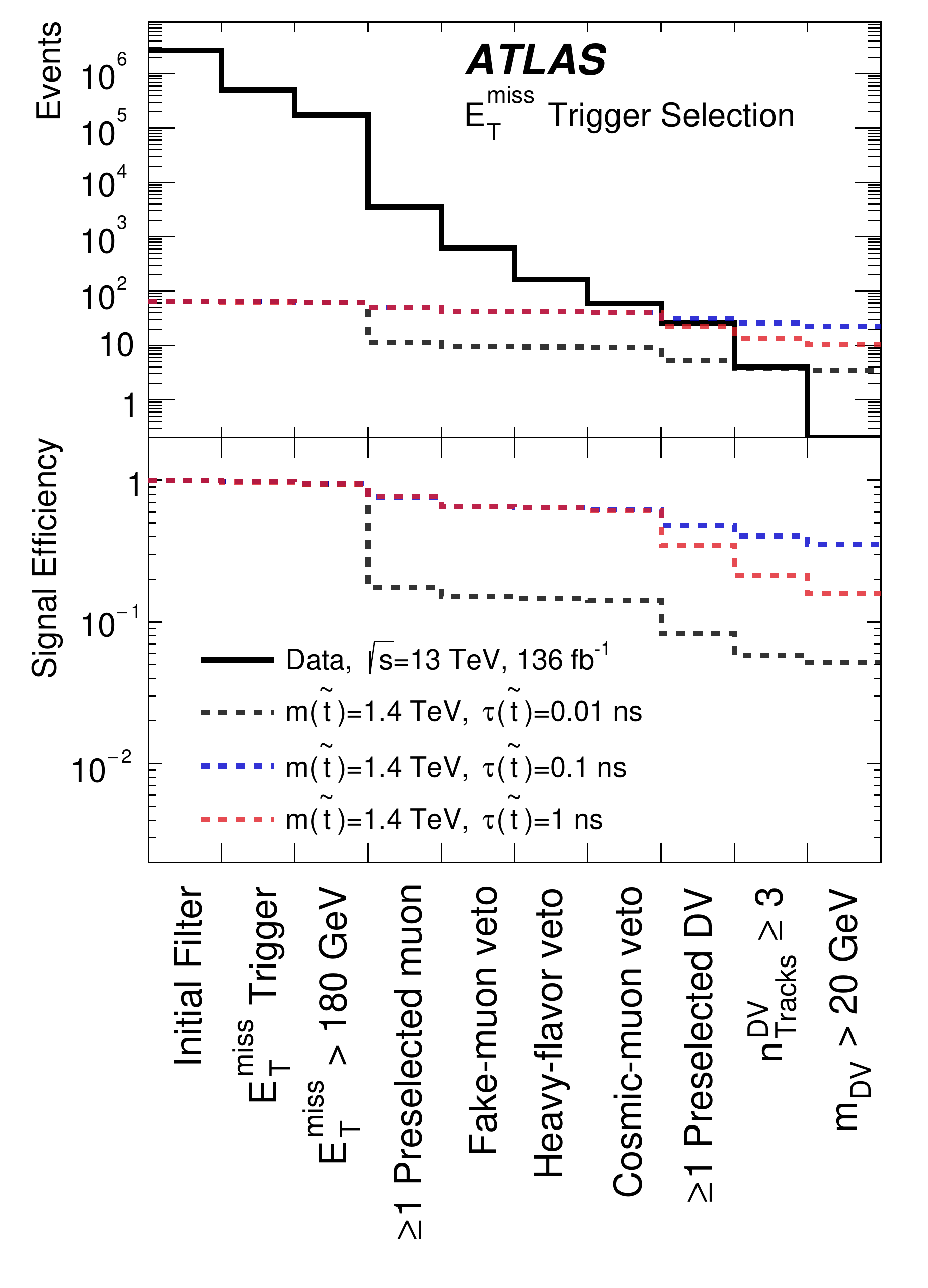}
\includegraphics[width=0.45\textwidth]{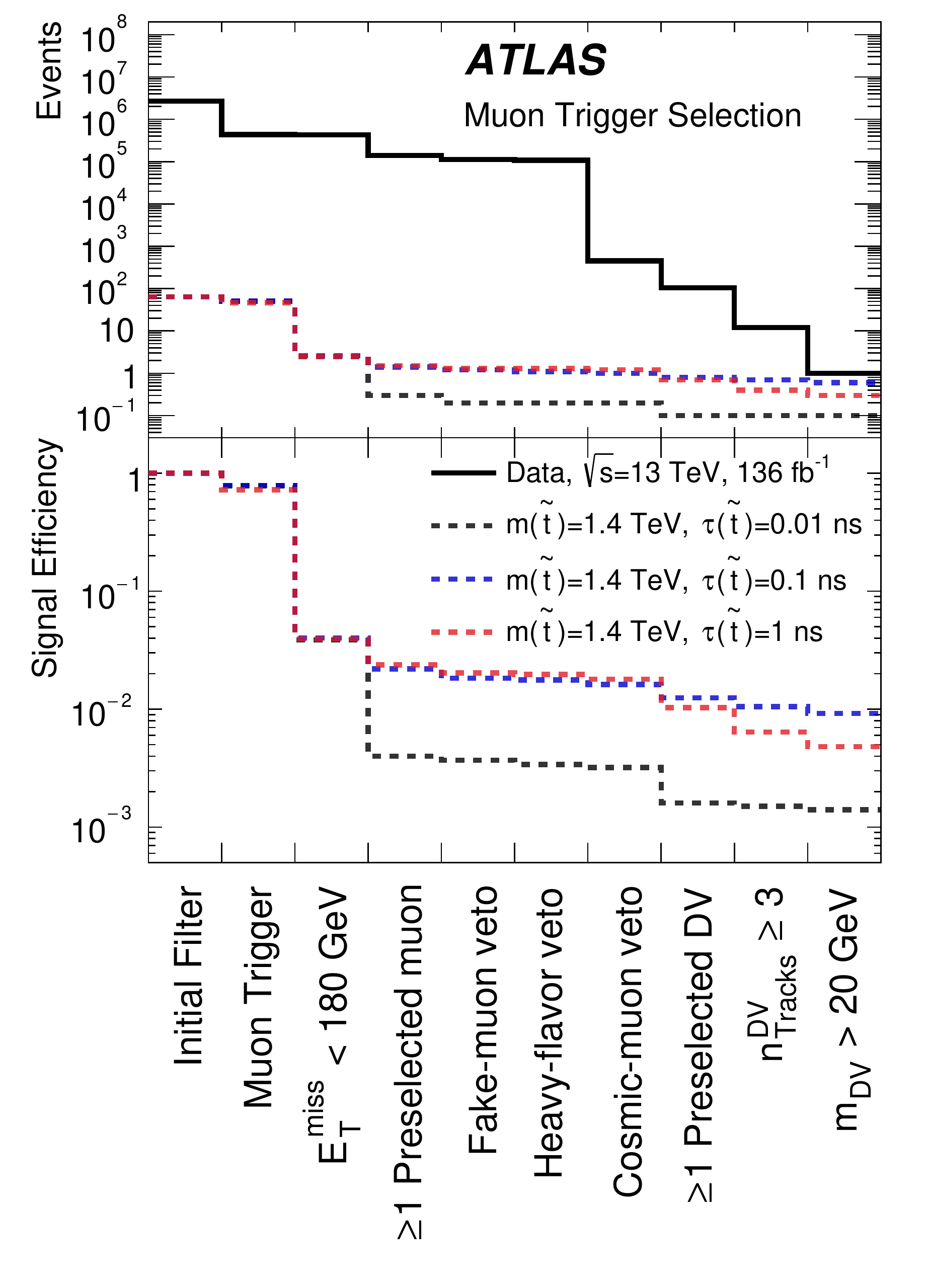}
\caption{Event yields are shown as a function of the SR requirements for data and three example signal models for the \met Trigger selection (left) and the Muon Trigger selection (right). The Initial Filter requirement includes selections on the data due to the raw-event filters. For the signals, no initial filter is applied and acceptance effects are included. The \met\ requirement label denotes requirements on the cluster-based \met. Signals with lifetimes of $0.01$~ns, $0.1$~ns, and $1$~ns are shown in black, blue, and red, respectively. Samples are shown for signals with $m(\stop)=1.4$~TeV. }
\label{fig:cutflow}
\end{center}
\end{figure}

Signal and data event yields are shown as a function of progressively stricter requirements in \Fig{\ref{fig:cutflow}} along with the cumulative selection efficiencies, including acceptance effects. For the signal model considered here, over $95\%$ of events have cluster-based $\met>180$~\GeV\ and consequently the $\MET$ trigger SR has the larger cumulative selection efficiency. For signals with $\tau(\stop)=0.1$~ns, the cumulative selection efficiency of the $\MET$ trigger SR is $35\%$. Signals with $\tau(\stop)=1$~ns have a cumulative selection efficiency of $15\%$, due to the reduced efficiency for reconstructed displaced tracks and displaced vertices at large radii. Signals with $\tau(\stop)=0.01$~ns have cumulative selection efficiencies between 5--6$\%$, because many $\stop$ squark decays occur within millimeters of the primary $pp$ vertex. The efficiency loss for signals with $\tau(\stop)=0.01$~ns is more pronounced due to the requirement of at least one muon with $|d_0|>2$~mm, and a DV which is at least 4~mm away from any $pp$ collision vertex in the event.
 
\begin{figure}[!tb]
\begin{center}
\includegraphics[width=0.45\textwidth]{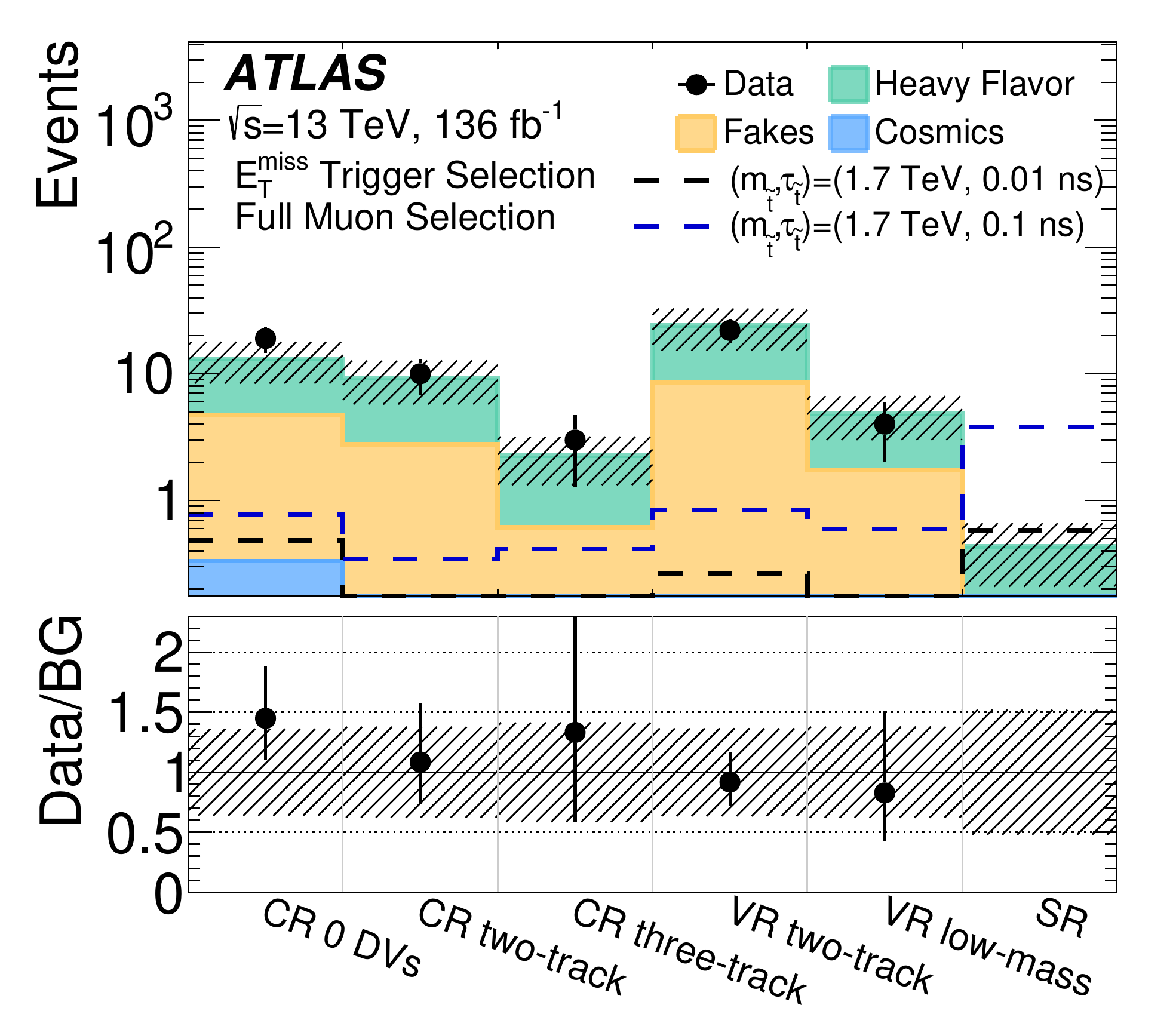}
\includegraphics[width=0.45\textwidth]{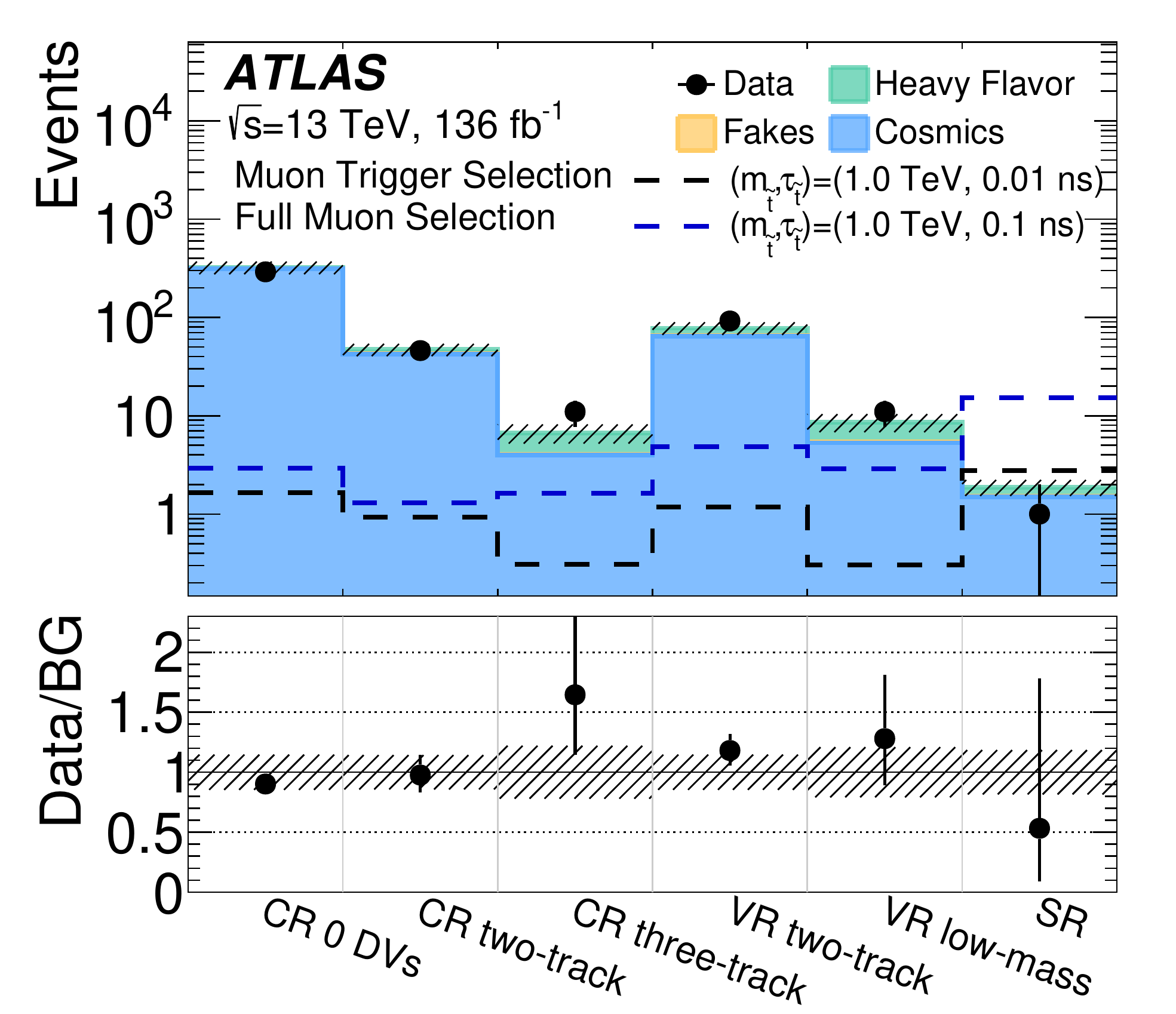}
\caption{The observed event yields in the control, validation and signal regions are shown for the \met\ Trigger (left) and Muon Trigger (right) selections, along with the predicted background yields. The bottom panel shows the ratio of observed events to the total background yields. Signals with lifetimes of $0.01$~ns and $0.1$~ns are shown in blue and red, respectively. For the $\met$-triggered (muon-triggered) selection, signal samples are shown with $m(\stop)=1.7$~TeV ($m(\stop)=1.0$~TeV). The errors represent the total uncertainty of the background predictions, including the statistical and systematic uncertainties added in quadrature.}
\label{fig:finalyields}
\end{center}
\end{figure}

The final yields in the control, validation and signal regions are shown in \Fig{\ref{fig:finalyields}}. In the DV VRs, the data show good agreement with the background predictions within uncertainties, validating the assumptions made in the background estimation. In the SRs, the data show good agreement with the background predictions within uncertainties.
 
\FloatBarrier

\begin{figure}[!tb]
\begin{center}
\includegraphics[width=0.75\textwidth]{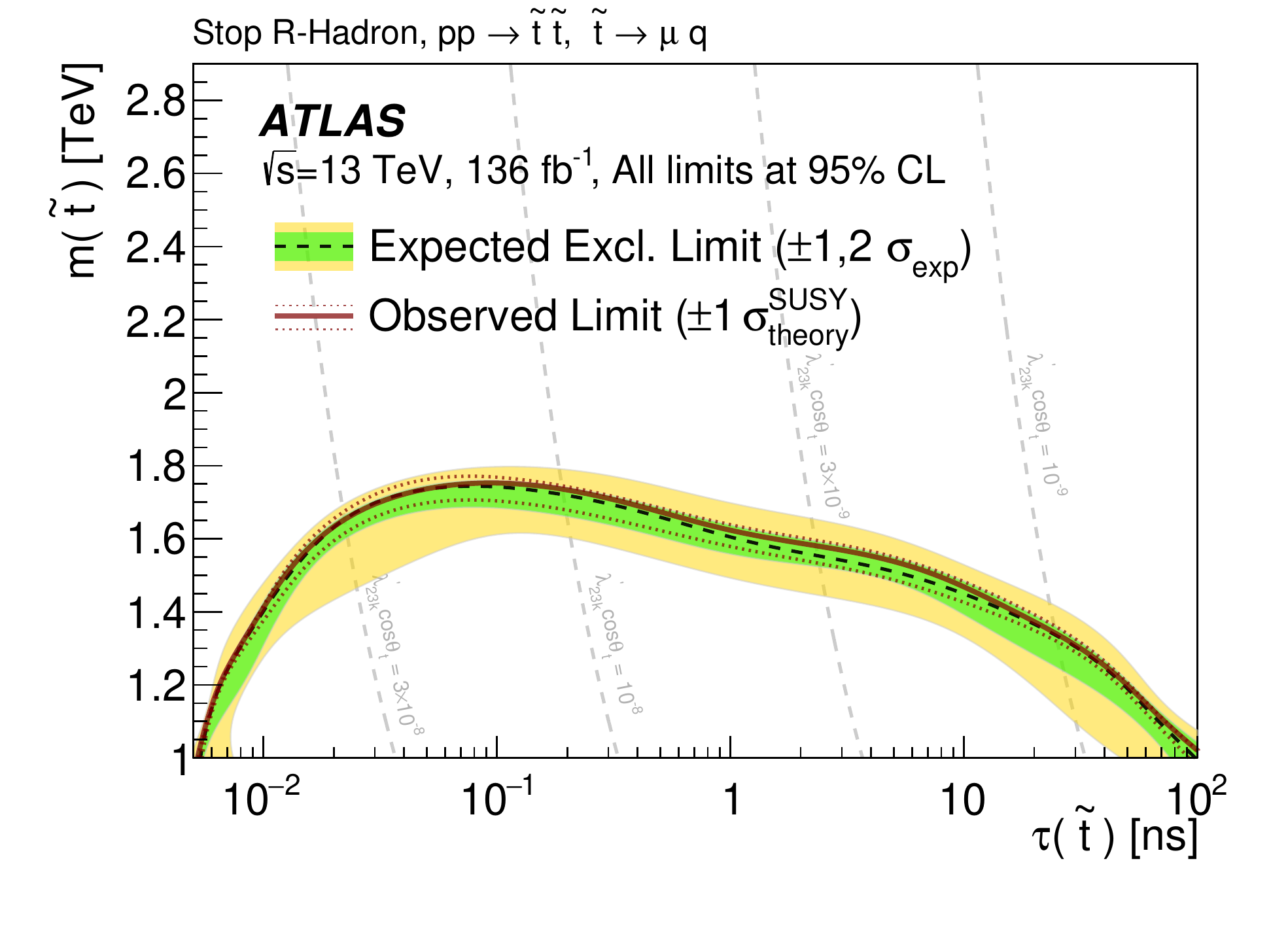} 
\caption{Exclusion limits at 95\% CL on $m(\stop)$ as a function of $\tau(\stop)$ are shown along with contours showing fixed values of $\lamp_{23k}\cos\theta_t$. Experimental and theoretical uncertainties are shown.}
\label{fig:modellimits_1}
\end{center}
\end{figure}

\begin{figure}[!tb]
\begin{center}
\includegraphics[width=0.75\textwidth]{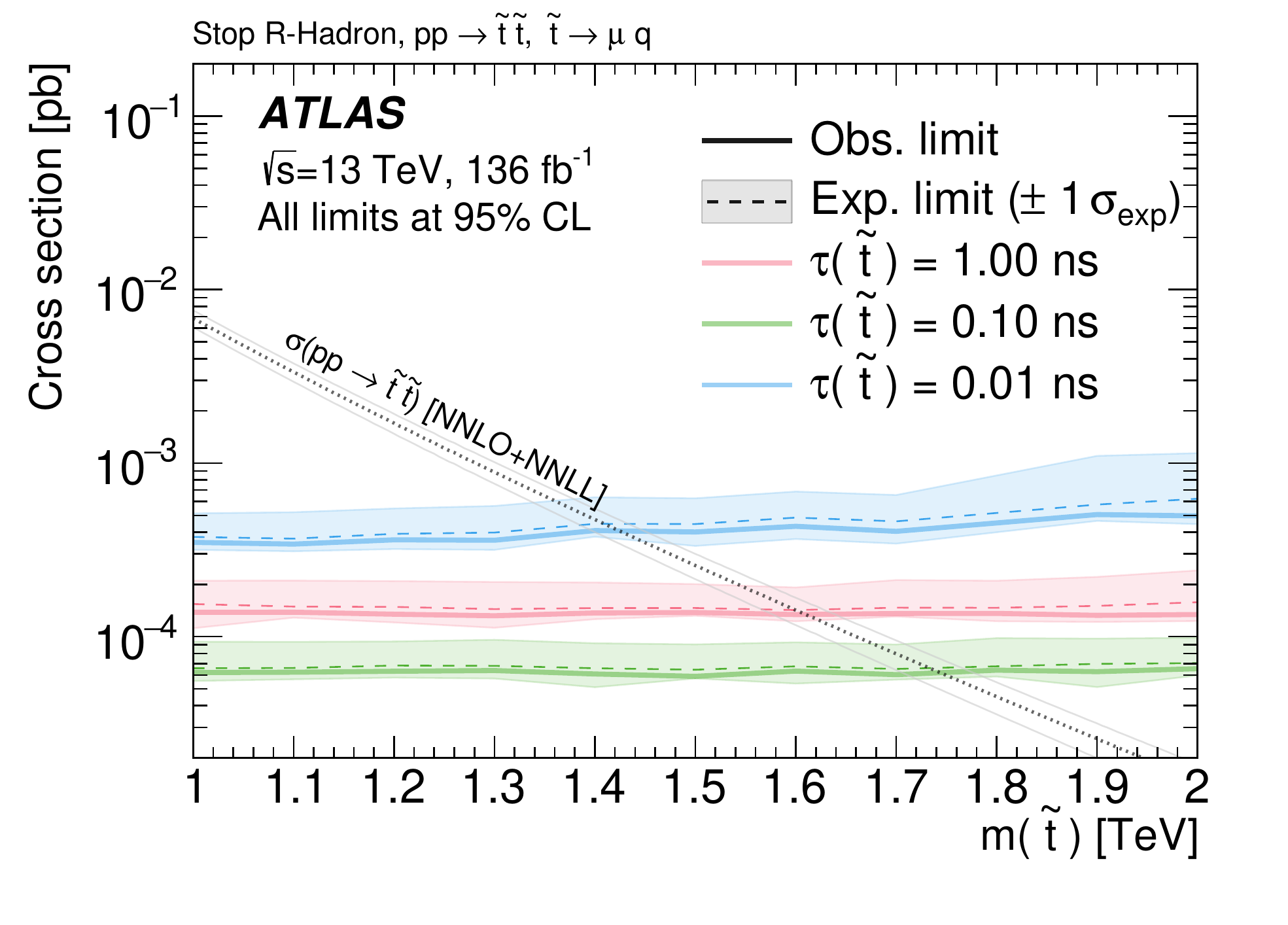}
\caption{Exclusion limits on the production cross section as a function of $m(\stop)$ are shown for several values of $\tau(\stop)$ along with the nominal signal production cross section and its theoretical uncertainty.}
\label{fig:modellimits_2}
\end{center}
\end{figure}

The \textsc{HistFitter} package~\cite{Baak:2014wma} was used for the calculation of the 95\% CL exclusion limits using the CL$_\mathrm{s}$ prescription~\cite{Read:2002hq}. \Fig{\ref{fig:modellimits_1}} shows expected and observed 95\% CL exclusion limits on the mass of a long-lived $\stop$ squark as a function of its mean proper lifetime $\tau(\stop)$. For each configuration of model parameters, the limit from the SR with the higher expected sensitivity is shown. The sensitivity to the shown region of parameter space comes entirely from the \met Trigger SR. Additionally, in \Fig{\ref{fig:modellimits_2}}, 95\% CL upper limits on the production cross section of these signals are shown as a function of $m(\stop)$ for various values of $\tau(\stop)$. The limits show little variation as a function of $m(\stop)$ owing to the fact that the signal acceptance and efficiency are relatively invariant in $m(\stop)$ for the range considered here.
 
Contours showing fixed values of $\lamp_{23k}\cos\theta_t$ are also shown in \Fig{\ref{fig:modellimits_1}} where $\theta_t$ is the mixing angle between the left- and right-handed \stop\ squarks.
These contours are derived using the expression provided in Ref.~\cite{Kon:1994xe} for a \stop\ squark decay via a single \lampijk\ coupling:
 
\begin{eqnarray*}
\tau(\stop) & = & \frac{16\pi}{m(\stop)~[\lambda'_{23k}  \cos \theta_t]^2 }
\end{eqnarray*}

For a mean proper lifetime of $\tau(\stop)=0.1$~ns, masses below roughly $m(\stop)=1.7$~\TeV\ are excluded. For masses on the order of the DV mass requirement of $20$~GeV and below, sensitivity is expected to drastically decrease. For the wide range of mean proper lifetime values between $0.01$~ns and $30$~ns, masses below $1.3$~\TeV\ are excluded at 95\% CL. For $\tau(\stop)=0.1$~ns, cross-section upper limits are set below $100$~ab. For mean proper lifetimes between $0.01$~ns and $100$~ns, these limits are the strictest to date on models with a metastable \stop\ squark decaying via the \lampijk\ RPV coupling. For $m(\stop)=1$~\TeV, values of $\lamp_{23k}\cos\theta_t$ between roughly $10^{-7}$ and $10^{-9}$ are excluded at 95\% CL.
 
Model-independent upper limits at 95\% CL on the number of BSM events in the signal region are also derived, assuming no significant contamination from alternate signal models in the control regions. Normalizing these limits by the integrated luminosity of the data sample, these numbers can be interpreted as upper limits on the visible BSM cross section, denoted by $\sigma_{\mathrm{vis}}$. It is defined as the product of signal acceptance, reconstruction efficiency, and production cross section. The results are given in \Tab{\ref{table.results.exclxsec.pval.upperlimit.SS}}.
 
\begin{table}
\caption[Breakdown of upper limits.]{The observed 95\% CL upper limits on the visible cross section
$\langle\sigma_{\mathrm{vis}}\rangle_{\textrm{obs} }^{95}$, and the observed ($S_{\textrm{obs} }^{95}$) and expected ($S_{\textrm{exp} }^{95}$) limits on the number of signal events.}
\centering
\setlength{\tabcolsep}{0.0pc}
\begin{tabular*}{0.7\textwidth}{@{\extracolsep{\fill}}lccccc}
\noalign{\smallskip}\hline\hline\noalign{\smallskip}
Signal Region                        &  $S_{\textrm{obs} }^{95}$  & $S_{\textrm{exp} }^{95}$ & $\langle\sigma_{\mathrm{vis}}\rangle_{\textrm{obs} }^{95}$~[fb] \\
\noalign{\smallskip}\hline\noalign{\smallskip}
\met Trigger SR    &  $3.0$ & $ { 3.0 }^{ +1.1 }_{ -0.0 }$ & $0.023$  \\%
Muon Trigger SR    &  $3.6$ & $ { 4.2 }^{ +1.7 }_{ -1.1 }$ & $0.027$ \\%
\noalign{\smallskip}\hline\hline\noalign{\smallskip}
\end{tabular*}
\label{table.results.exclxsec.pval.upperlimit.SS}
\end{table}

\section{Conclusion}
\label{sec:conclusion}
 
A search for physics beyond the Standard Model giving rise to long-lived particle decays with muons is performed with the ATLAS experiment at the LHC using $\intlumi$ of $pp$ collision data at $\sqrt{s} = 13$~\TeV. Event selections are developed to efficiently reject backgrounds. The yields expected from background in the two orthogonal signal regions used in the analysis are extracted from control regions in the data and amount to \metSRYield\ in the \met Trigger SR, and \muonSRYield\ in the Muon Trigger SR. The data agree with the yields expected from the background-only hypothesis, with zero and one event passing the \met and Muon Trigger SR requirements, respectively.
 
The results are interpreted in a supersymmetric model with pair-produced top squarks \stop\ decaying via small values of the $R$-parity-violating coupling $\lamp_{23k}$ into a muon and a quark, giving the \stop\ squark mean proper lifetimes $\tau(\stop)$ in the ps to ns range. At 95\% confidence level, $m(\stop)$ values up to 1.7~\TeV\ for $\tau(\stop) = 0.1$~ns are excluded, and the limit surpasses 1.3~\TeV\ for all lifetimes in the range from $0.01$~ns to $30$~ns. Upper limits on the visible cross section for processes with the sought signature are 0.023~fb and 0.027~fb for the $\met$ and Muon Trigger SRs, respectively.
 
\section*{Acknowledgements}
 

We thank CERN for the very successful operation of the LHC, as well as the
support staff from our institutions without whom ATLAS could not be
operated efficiently.
 
We acknowledge the support of ANPCyT, Argentina; YerPhI, Armenia; ARC, Australia; BMWFW and FWF, Austria; ANAS, Azerbaijan; SSTC, Belarus; CNPq and FAPESP, Brazil; NSERC, NRC and CFI, Canada; CERN; CONICYT, Chile; CAS, MOST and NSFC, China; COLCIENCIAS, Colombia; MSMT CR, MPO CR and VSC CR, Czech Republic; DNRF and DNSRC, Denmark; IN2P3-CNRS and CEA-DRF/IRFU, France; SRNSFG, Georgia; BMBF, HGF and MPG, Germany; GSRT, Greece; RGC and Hong Kong SAR, China; ISF and Benoziyo Center, Israel; INFN, Italy; MEXT and JSPS, Japan; CNRST, Morocco; NWO, Netherlands; RCN, Norway; MNiSW and NCN, Poland; FCT, Portugal; MNE/IFA, Romania; MES of Russia and NRC KI, Russia Federation; JINR; MESTD, Serbia; MSSR, Slovakia; ARRS and MIZ\v{S}, Slovenia; DST/NRF, South Africa; MINECO, Spain; SRC and Wallenberg Foundation, Sweden; SERI, SNSF and Cantons of Bern and Geneva, Switzerland; MOST, Taiwan; TAEK, Turkey; STFC, United Kingdom; DOE and NSF, United States of America. In addition, individual groups and members have received support from BCKDF, CANARIE, Compute Canada and CRC, Canada; ERC, ERDF, Horizon 2020, Marie Sk{\l}odowska-Curie Actions and COST, European Union; Investissements d'Avenir Labex, Investissements d'Avenir Idex and ANR, France; DFG and AvH Foundation, Germany; Herakleitos, Thales and Aristeia programmes co-financed by EU-ESF and the Greek NSRF, Greece; BSF-NSF and GIF, Israel; CERCA Programme Generalitat de Catalunya and PROMETEO Programme Generalitat Valenciana, Spain; G\"{o}ran Gustafssons Stiftelse, Sweden; The Royal Society and Leverhulme Trust, United Kingdom.
 
The crucial computing support from all WLCG partners is acknowledged gratefully, in particular from CERN, the ATLAS Tier-1 facilities at TRIUMF (Canada), NDGF (Denmark, Norway, Sweden), CC-IN2P3 (France), KIT/GridKA (Germany), INFN-CNAF (Italy), NL-T1 (Netherlands), PIC (Spain), ASGC (Taiwan), RAL (UK) and BNL (USA), the Tier-2 facilities worldwide and large non-WLCG resource providers. Major contributors of computing resources are listed in Ref.~\cite{ATL-GEN-PUB-2016-002}.
 

\appendix

\clearpage
 
\printbibliography

\clearpage 
 
\begin{flushleft}
\hypersetup{urlcolor=black}
{\Large The ATLAS Collaboration}

\bigskip

\AtlasOrcid[0000-0002-6665-4934]{G.~Aad}$^\textrm{\scriptsize 102}$,    
\AtlasOrcid[0000-0002-5888-2734]{B.~Abbott}$^\textrm{\scriptsize 129}$,    
\AtlasOrcid{D.C.~Abbott}$^\textrm{\scriptsize 103}$,    
\AtlasOrcid[0000-0002-2788-3822]{A.~Abed~Abud}$^\textrm{\scriptsize 71a,71b}$,    
\AtlasOrcid[0000-0002-1002-1652]{K.~Abeling}$^\textrm{\scriptsize 53}$,    
\AtlasOrcid[0000-0002-2987-4006]{D.K.~Abhayasinghe}$^\textrm{\scriptsize 94}$,    
\AtlasOrcid[0000-0002-8496-9294]{S.H.~Abidi}$^\textrm{\scriptsize 167}$,    
\AtlasOrcid[0000-0002-8279-9324]{O.S.~AbouZeid}$^\textrm{\scriptsize 40}$,    
\AtlasOrcid{N.L.~Abraham}$^\textrm{\scriptsize 156}$,    
\AtlasOrcid[0000-0001-5329-6640]{H.~Abramowicz}$^\textrm{\scriptsize 161}$
% ...
\bigskip
\\

% ...
$^{40}$Niels Bohr Institute, University of Copenhagen, Copenhagen; Denmark.\\
% ...
$^{53}$II. Physikalisches Institut, Georg-August-Universit\"{a}t G\"ottingen, G\"ottingen; Germany.\\
% ...
$^{71}$$^{(a)}$INFN Sezione di Pavia;$^{(b)}$Dipartimento di Fisica, Universit\`a di Pavia, Pavia; Italy.\\
% ...
$^{94}$Department of Physics, Royal Holloway University of London, Egham; United Kingdom.\\
% ...
$^{102}$CPPM, Aix-Marseille Universit\'e, CNRS/IN2P3, Marseille; France.\\
$^{103}$Department of Physics, University of Massachusetts, Amherst MA; United States of America.\\
% ...
$^{129}$Homer L. Dodge Department of Physics and Astronomy, University of Oklahoma, Norman OK; United States of America.\\
% ...
$^{156}$Department of Physics and Astronomy, University of Sussex, Brighton; United Kingdom.\\
% ...
$^{161}$Raymond and Beverly Sackler School of Physics and Astronomy, Tel Aviv University, Tel Aviv; Israel.\\
% ...
$^{167}$Department of Physics, University of Toronto, Toronto ON; Canada.\\
% ...
$^{a}$ Also at Borough of Manhattan Community College, City University of New York, New York NY; United States of America.\\
$^{b}$ Also at CERN, Geneva; Switzerland.\\
% ...

\end{flushleft}


\end{document}